\documentclass[fleqn,usenatbib]{mnras}

\usepackage{newtxtext,newtxmath}

\usepackage{gensymb}
\usepackage{mathtools}
\usepackage{threeparttable}
\usepackage{siunitx}
\usepackage[T1]{fontenc}

\DeclareRobustCommand{\VAN}[3]{#2}
\let\VANthebibliography\thebibliography
\def\thebibliography{\DeclareRobustCommand{\VAN}[3]{##3}\VANthebibliography}

\usepackage{graphicx}	
\usepackage{amsmath}	
\usepackage{xcolor}

\title[RS Oph with LOFAR and MeerKAT]{Low-frequency radio observations of recurrent nova RS Ophiuchi with MeerKAT and LOFAR}

\author[I. de Ruiter et al.]{
Iris de Ruiter$^{1}$\thanks{E-mail: i.deruiter@uva.nl},
Miriam M. Nyamai$^{2}$,
Antonia Rowlinson$^{1,3}$,
Ralph A.M.J. Wijers$^{1}$,
Tim J. O'Brien$^{4}$,
\newauthor
David R. A. Williams$^{4}$,
and Patrick Woudt$^{5}$
\\
$^{1}$Anton Pannekoek Institute for Astronomy, University of Amsterdam, Science Park 904, NL-1098 XH Amsterdam, the Netherlands\\
$^{2}$ South African Radio Observatory (SARAO), 2 Fir Street, Black River Park, Observatory, Cape Town, 7925, South Africa\\
$^{3}$ASTRON, the Netherlands Institute for Radio Astronomy, Postbus 2, NL-7990 AA Dwingeloo, the Netherlands\\
$^{4}$Jodrell Bank Centre for Astrophysics, School of Physics and Astronomy, University of Manchester, Manchester M13 9PL, UK\\
$^{5}$Department of Astronomy, University of Cape Town, Private Bag X3, Rondebosch 7701, South Africa
}

\date{Accepted XXX. Received YYY; in original form ZZZ}

\pubyear{2023}

\begin{document}
\label{firstpage}
\pagerange{\pageref{firstpage}--\pageref{lastpage}}
\maketitle

\begin{abstract}
We report low-frequency radio observations of the 2021 outburst of the recurrent nova RS Ophiuchi. These observations include the lowest frequency observations of this system to date. Detailed light curves are obtained by MeerKAT at 0.82 and 1.28 GHz and LOFAR at 54 and 154 MHz. These low-frequency detections allow us to put stringent constraints on the brightness temperature that clearly favour a non-thermal emission mechanism. The radio emission is interpreted and modelled as synchrotron emission from the shock interaction between the nova ejecta and the circumbinary medium. The light curve shows a plateauing behaviour after the first peak, which can be explained by either a non-uniform density of the circumbinary medium or a second emission component. Allowing for a second component in the light curve modelling captures the steep decay at late times. Furthermore, extrapolating this model to 15 years after the outburst shows that the radio emission might not fully disappear between outbursts. Further modelling of the light curves indicates a red giant mass loss rate of $\sim 5 \cdot 10^{-8}~{\rm M_\odot~yr^{-1}}$. The spectrum cannot be modelled in detail at this stage, as there are likely at least four emission components. Radio emission from stellar wind or synchrotron jets is ruled out as the possible origin of the radio emission. Finally, we suggest a strategy for future observations that would advance our understanding of the physical properties of RS Ophiuchi.

\end{abstract}





\section{Introduction}

The recurrent nova RS Ophiuchi (RS Oph) was discovered to be in outburst on $2021$ August $8$\footnote{AAVSO Alert Notice 752 by K. Geary/VSNET alert N.26131}. RS Oph is one of the best-studied members of the small sample of recurrent nova systems (see \cite{anupama2008recurrent} for an overview and \cite{bode2008classical, darnley2021accrete} for a review). RS Oph is a binary system, consisting of a hot, accreting white dwarf and a red giant companion, which qualifies the system as a symbiotic binary. The system has an orbital period of $454$ days \citep{brandi2009spectroscopic, dobrzycka1994new, fekel2000infrared}. 
The white dwarf accretes matter from the red giant either via a Roche-Lobe-filling secondary \citep{brandi2009spectroscopic, booth2016modelling} or the stellar wind of the secondary \citep{starrfield2008rs, wynn2008accretion}. Approximately every $20$ years the system goes into outburst as a nova. These outbursts are caused by a thermonuclear explosion in the hydrogen-rich white dwarf envelope, following accretion from the secondary \citep{starrfield1985recurrent}. The nova explosion expels accreted material into the circumstellar medium and significantly brightens the star \citep{bode2008classical}. Novae are usually detected in the optical,  triggering a multi-wavelength follow-up in order to study both the interaction of the nova ejecta with the environment and the post-thermonuclear runaway nuclear burning on the white dwarf. Previously recorded outbursts of RS Oph occurred in irregular intervals ($1898$, $1933$, $1958$, $1967$, $1985$, $2006$) with two more candidates suggested in the literature ($1907$; \cite{schaefer2004rs} and $1945$; \citealt{oppenheimer1993analysis}). The frequency of these outbursts implies a high accretion rate and suggests that the white dwarf mass is close to the Chandrasekhar limit \citep{hachisu2001recurrent, sokoloski2006x}, making RS Oph a compelling candidate Type Ia supernova progenitor. A detailed understanding of RS Oph and its environment thus furthers our understanding of stellar evolution.

The $2021$ outburst of RS Oph has been monitored using a wide variety of facilities. Swiftly after the optical discovery, telescopes over the full electromagnetic spectrum started monitoring the system. The $2021$ outburst was promptly discovered in high-energy ($0.1$-$10$ GeV) gamma-rays \citep{cheung2022fermi} by \textit{Fermi-LAT}. Very high energy gamma-rays were detected by MAGIC \citep{acciari2022gamma} and H.E.S.S. \citep{hess2022time}. X-ray emission was detected by multiple facilities including \textit{NICER} \citep{ATEL_nicer}, \textit{INTEGRAL} \citep{ATEL_integral} and the \textit{Neil Gehrels Swift Observatory} \citep{page2022swift}. Radio observations have been made with AMI-LA, \textit{e}-MERLIN, MeerKAT \citep{ATEL_ami}, the VLA \citep{ATEL_vla}, VLITE \citep{ATEL_vlite} and LOFAR. Naturally, the outburst was monitored by a variety of optical facilities. Finally, follow-up with IceCube placed upper limits on the neutrino emission \citep{ATEL_icecube}.

\begin{figure}
    \centering
    \includegraphics[width=\linewidth]{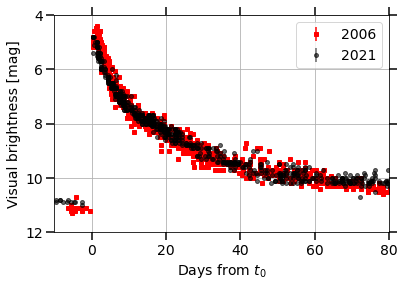}
    \caption{Visual band photometry from the AAVSO Collaboration, from both $2021$ and $2006$ \protect\citep{kafka2016observations}. The two outbursts are almost identical at optical wavelengths. For $2006$ the JD assumed for $t_0=2453779.33$ and for $2021$ $t_0=2459435$.}
    \label{fig:optical_lightcurve}
\end{figure}

The optical light curves of recurrent novae decay quickly compared to many classical novae, due to the lower amount of material accreted onto the white dwarf surface since the previous outburst. Figure \ref{fig:optical_lightcurve} shows the optical light curve of RS Oph as obtained by the AASVO for the $2021$ outburst (black dots) and the $2006$ outburst (red squares). The two outbursts are almost identical at optical wavelengths, indicating a similar white dwarf mass, ejecta mass, ejecta velocity and geometry \citep{shore2012spectroscopy, chomiuk2021new}. Throughout this work we will often compare the $2021$ outburst to the $2006$ outburst.

The radio light curves of novae typically show an initial rise, during which the ejecta are (at least partially) optically thick, with a positive spectral index, followed by an optically thin decline. Throughout this work, we define the spectral index as $\alpha$ where $S_{\nu} \propto \nu^{\alpha}$. Studies of the spectral development of the $2006$ outburst found that there are likely two emission mechanisms; a fading non-thermal emission subject to variable absorption, dominating at low frequencies \citep{kantharia2007giant} and an additional thermal component, dominating at higher frequencies \citep{eyres2009double}. This multi-component interpretation is consistent with high-resolution long-baseline images at higher frequencies \citep{o2006asymmetric, rupen2008expanding, sokoloski2008uncovering}.

Possible origins of the radio emission are synchrotron radiation produced by relativistic electrons in a magnetic field or thermal bremsstrahlung from ionized (hydrogen) gas. 
Evidence of non-thermal emission from RS Oph includes high brightness temperatures and an absence of X-ray flux compared to the observed radio flux in a thermal scenario for the $1985$ outburst \citep{taylor1989vlbi}. In the $2006$ outburst, a spectral index of $\alpha \sim -0.8$ indicated a clearly non-thermal origin of the radio emission \citep{kantharia2007giant} for observations in the range of 0.24 to 1.46 GHz. 
In this scenario, the nova ejecta shock the circumbinary red giant material, which is heated and ionized by the nova, generating relativistic electrons and enhanced magnetic fields, necessary for synchrotron emission \citep{chevalier1982radio, chevalier1982young}. Synchrotron radiation dominates in other recurrent novae embedded in red giant winds, see for example V3890 Sgr \citep{nyamai2021radio}, V745 Sco \citep{kantharia2015insights} and V1535 Sco \citep{linford2017peculiar}.

In this paper, we present low-frequency radio observations of RS Oph using the LOw-Frequency ARray \citep[LOFAR;][]{van2013lofar} and the Meer Karoo Array Telescope \citep[MeerKAT;][]{jonas2009meerkat}. We report the detection, for the first time, of RS Oph at frequencies below $240$ MHz and present a unique high-cadence light curve at MeerKAT frequencies. We present our observations and data reduction methods in Section \ref{sec:observations}. The light curves and spectrum resulting from these observations are presented in Section \ref{sec:results}. Section \ref{sec:discussion} describes the physics one can derive from these low-frequency radio observations by calculating the brightness temperature, modelling of the light curve and emission components and determining the equipartition magnetic field strength. We discuss why stellar winds and synchrotron jets can be ruled out as possible sources of the observed radio emission in Section \ref{sec:alternative_scenarios}. Finally, we discuss future observing strategies in Section \ref{sec:future} and conclude and summarize in Section \ref{sec:conclusion}. 

\section{Observations} 
\label{sec:observations} 

\subsection{MeerKAT observations}
During its $2021$ outburst, RS Oph was observed with MeerKAT at $0.82$ GHz (UHF band) and $1.28$ GHz (L band). The monitoring of the nova started on $t-t_0~ =~ 1$ day where $t_0$ is taken as $2021$ August $08.5$ (MJD $59434.5$) \citep{munari20212021}. 
For the L-band receiver, the frequency coverage is between $0.9$ to $1.67$ GHz centred at $1.284$ GHz. For the UHF band receiver, the frequency coverage is between $0.58$ to $1.015$ GHz centred at $0.816$ GHz. For each epoch, the time on target was either $15$ minutes or $30$ minutes (see Table \ref{tab:obs}). For the first $10$ days following the discovery of the nova in the optical wavelengths, RS Oph was monitored daily by MeerKAT. Afterwards, the cadence was reduced and the last observation of the nova was obtained on day $223$.

RS Oph was bracketed by $2$ minute phase calibrator scans on the source J1733-1304 to solve for complex gains. A bandpass calibrator (J1939-6342) was observed for $5-10$ minutes at the beginning of each observation. Bandpass corrections were determined for the primary calibrator and complex gains were determined for both primary and secondary calibrators. The corrections from the primary calibrator were used to determine the absolute flux density of the secondary calibrator. The calibration solutions and absolute flux density scale were then transferred to the target (RS Oph). The calibration and imaging steps followed in data reduction are summarised in the {\sc oxkat} pipeline\footnote{https://www.
github.com/IanHeywood/oxkat and through the Astrophysics Source Code Library record ascl:2009.003 \citep{heywood2020oxkat}.}.

Analysis of the MeerKAT results was performed using CASA \citep{bean2022casa}. The flux densities  and errors for each epoch were determined using the \texttt{\textbf{imfit}} task within the target region. RS Oph was unresolved at both observing frequency bands with a telescope resolution (FWHM) of $12"$ and $8"$ at $1.284$ GHz and $0.816$ GHz, respectively. The results are presented in Table \ref{tab:obs} and Figure \ref{fig:lightcurve}. The quoted uncertainties include the MeerKAT flux density calibration accuracy of $10\%$ and Gaussian fit error from \texttt{\textbf{imfit}} (added in quadrature, see Eqn.\ref{eq:uncertannity}).

\subsection{LOFAR HBA}
RS Oph was first observed with the high band antennas (HBA) at 154 MHz on day $25$ after optical discovery. These observations were the result of DDT proposals (\texttt{DDT16\_001} and \texttt{DDT17\_002}) that were prompted by the early MeerKAT detections. The LOFAR observations were scheduled to be simultaneous with MeerKAT, \textit{e}-MERLIN and EVN, with observation dates as shown in Table \ref{tab:obs}. Each observation had a duration of $2$ hours. The calibrator 3C295 was observed for $10$ minutes after each observation. The HBA observations were obtained with the full Dutch array and array configuration HBA DUAL INNER \citep{van2013lofar}. 
Calibration was performed with  \textsc{prefactor}\footnote{\url{https://github.com/lofar-astron/prefactor}} and a strategy based upon that presented in \cite{van2016lofar}. During calibration, the instrumental and ionospheric effects present in the LOFAR data are corrected by iteratively comparing observations of calibrators with a LOFAR sky model of the calibrator sources. This yields direction-independent gain corrections that are applied to the target observations. The final images have an rms noise between $1.9$ and $3.1$ mJy/beam, as measured in a box well away from the target location.

Both the target and calibrator observations were flagged for excess radio frequency interference using \textsc{AOFlagger} \citep{offringa2010post, offringa2012morphological}. 
We imaged the full LOFAR observation using \textsc{WSClean} \citep{offringa2014wsclean} using a primary beam correction, Briggs weighting with a robustness of $-0.5$, a pixel scale of $5$ arcsec, and baselines up to $8$ k$\lambda$. Cleaning was conducted using $100000$ iterations. The final image has a central frequency of $154$ MHz and a bandwidth of $48$ MHz.  Using the Python Source Extractor \citep[{\sc PySE};][]{carbone2018pyse}, a flux density was extracted at the position of RS Oph holding the shape and size of the Gaussian fitted fixed to the restoring beam shape, as RS Oph is a point source at these frequencies. The resolution of these observations is around $3$ arcsec \citep{van2013lofar}, which is too low to resolve the $\sim 22$ mas structures seen at higher frequencies, e.g., \cite{munari2022radio}. The integrated flux densities for each observation are given in Table \ref{tab:obs}. The flux errors in this table consist of a fit error $\sigma_{\rm{fit}}$, determined by \textsc{PySE}, and a $10\%$ systematic flux error \citep{shimwell2022lofar}. The total quoted error in Table \ref{tab:obs} is thus defined as:
\begin{equation}
    \sigma_{S_{\nu}} = \sqrt{\sigma_{\rm{fit}}^2 + (0.1 S_{\nu})^2}.
    \label{eq:uncertannity}
\end{equation}

\subsection{LOFAR LBA}

Since RS Oph was detected with high significance in the first HBA observation, a LOFAR low band antenna (LBA) observation at 54 MHz was made on day $49$ (see Table \ref{tab:obs}). The LBA observation was conducted immediately before the third HBA observation and had a duration of two hours, during which one sub-array pointing was centered on the position of RS Oph and one sub-array pointing was centered on the calibrator 3C295. 

The data were processed using the Library for Low Frequencies tools (\textsc{LiLF}\footnote{\url{https://github.com/revoltek/LiLF/tree/LBAdevel}}). In summary, the steps consist of reducing the calibrator data \citep{de2019systematic}, applying these calibrator solutions to the target field and additional calibration to correct for differential ionospheric effects \citep{de2020reaching} and performing facet based direction dependent calibration with \textsc{DDFacet} \citep{tasse2018faceting}. The first steps of calibration are similar to the ones described in the previous section. However, additional direction-dependent corrections are applied via self-calibration. Using just the direction-independent calibration from the calibrator may leave residual errors in the data because the calibrator sources are observed at a different position on the sky than the target sources. These residual errors can be especially large for LBA observations, which are severely affected by the ionosphere \citep{de2020reaching}. In the process of self-calibration, the sky model is updated using the target field and this sky model is used to recalibrate the data. This process is repeated until no major improvements are made on the image quality.

After calibration, a forced source extraction was performed at the position of RS Oph in an image of one of the calibrators after direction dependent calibration. The observation on day $49$ yielded a detection at $3.5$ x rms noise, after excluding $3$ LOFAR stations with poor amplitude solutions. The rms noise in the final image is $16.9$ mJy/beam. The flux density value obtained by forced flux extraction of a beam-sized point source at the location of RS Oph is quoted in Table \ref{tab:obs}. The error is again calculated by combining the fit error and a systematic error of 10\% \citep{de2021lofar}, where for the LBA image the fit error dominates. 

The marginal detection at day $49$ indicated that ideally the LBA observation would be of longer duration because extensive flagging at low frequencies drastically reduces the useful amount of data. Therefore, the second LBA observation was split over three days and each observation had a duration of $2$ hours, making the total observing time 6 hours. This second LBA observation yielded an image with an rms noise level of $7.42$ mJy/beam. RS Oph was not detected in this second LBA observation. The $3$ × rms upper limit for RS Oph during this observation is $22.26$ mJy. Nevertheless, a forced flux extraction was performed, resulting in a flux density with extremely large uncertainty, as presented in Table \ref{tab:obs}.

\begin{table*}
    \centering
    \caption{Flux density values from observations with MeerKAT ($1284$ MHz and $816$ MHz) and LOFAR HBA (154 MHz) and LBA (54 MHz). Here $t_0$ is the start of the nova eruption $2021$ August $08.5$ (MJD $59434.5$)  \citep{munari20212021}. The observation time is the on-source observation time without calibrator observations, except for the LOFAR LBA observations where target and calibrator observations are performed simultaneously.}
    \label{tab:obs}

\begin{threeparttable}
\begin{tabular}{l r@{}l r@{}l  r@{}l cc r@{}l c r@{}l}

\hline 
  &  & & \multicolumn{2}{c}{}  & \multicolumn{2}{c}{Observation}  &  & Observation &  \multicolumn{5}{c}{} \\
Start date - time  & $t$ & & \multicolumn{2}{c}{($t-t_0$)}  & \multicolumn{2}{c}{frequency} & Telescope &  time &  \multicolumn{5}{c}{$S_{\nu}$}  \\
(UTC) & (MJD) &  & \multicolumn{2}{c}{(Days)}  & \multicolumn{2}{c}{(MHz)} &  & (mins) & \multicolumn{5}{c}{(mJy)}  \\ 
\hline
2021-08-10.95 &  59437 & .0     & 2 & .5   & 1&284  & MeerKAT   & 30    & 0 & .35  & $\pm$ & 0 & .048 \\
2021-08-11.71 &  59437 & .7     & 3 & .2   & 1&284  & MeerKAT   & 15    & 0 & .28  & $\pm$ & 0 & .045 \\
2021-08-11.75 &  59437 & .8     & 3 & .3   & &816   & MeerKAT   & 30    & 0 & .49  & $\pm$ & 0 & .068 \\
2021-08-12.63 & 59438 & .6      & 4 & .1   & 1&284  & MeerKAT   & 15    & 0 & .59  & $\pm$ & 0 & .087\\
2021-08-13.61 & 59439 & .6      & 5 & .1   & 1&284  & MeerKAT   & 15    & 1 & .7  & $\pm$ & 0 & .18 \\
2021-08-14.58 & 59440 & .6      & 6 & .1   & &816   & MeerKAT   & 15    & 1 & .4  & $\pm$ & 0 & .16 \\
2021-08.14.61 & 59440 & .6      & 6 & .1   & 1&284  & MeerKAT   & 15    & 8 & .6  & $\pm$ & 0 & .87\\
2021-08-15.76 & 59441 & .8      & 7 & .3   & &816   & MeerKAT   & 15    &  9 & .8  & $\pm$ & 1 & .0 \\
2021-08-15.80 & 59441 & .8      & 7 & .3   & 1&284  & MeerKAT   & 15    & 32 & .9  & $\pm$ & 3 & .3 \\
2021-08-16.77 & 59442 & .8      & 8 & .3   & 1&284  & MeerKAT   & 15    & 52 & .4  & $\pm$ & 5 & .3\\
2021-08-17.60 & 59443 & .6      & 9 & .1   & 1&284  & MeerKAT   & 15    & 64 & .5  & $\pm$ & 6 & .5 \\
2021-08-18.82 & 59444 & .8      & 10 & .3  & &816   & MeerKAT   & 15    &  54 & .5  & $\pm$ & 5 & .5 \\
2021-08-18.86 & 59444 & .9      & 10 & .4  & 1&284  & MeerKAT   & 15    & 70 & .5  & $\pm$ & 7 & .1 \\
2021-08-23.72 & 59449 & .7      & 15 & .2  & 1&284  & MeerKAT   & 15    & 87 & .4  & $\pm$ & 8 & .7 \\
2021-08-23.76 & 59449 & .8      & 15 & .3  & &816   & MeerKAT   & 15    & 85 & .0  & $\pm$ & 8 & .5 \\
2021-09-01.64& 59458 & .6       & 24 & .1  & 1&284  & MeerKAT   & 15    & 76 & .1  & $\pm$ & 7 & .6 \\
2021-09-01.64 & 59458 & .6      & 24 & .1  & &816   & MeerKAT   & 15    & 78 & .7  & $\pm$ & 7 & .9\\
2021-09-01.74 &  59458 & .7     & 24 & .2  & &154   & LOFAR     & 120   & 39 & .9  & $\pm$ & 5 & .7   \\
2021-09-11.63 & 59468 & .6      & 34 & .1  & &816   & MeerKAT   & 15    & 73 & .5  & $\pm$ & 7 & .4  \\
2021-09-11.67 & 59468 & .7      & 34 & .2  & 1&284  & MeerKAT   & 15    & 70 & .8  & $\pm$ & 7 & .1  \\
2021-09-11.71 & 59468 & .7      & 34 & .2  & &154   & LOFAR     & 120   & 64 & .4  & $\pm$ & 7 & .6  \\
2021-09-26.59 & 59483 & .6      & 49 & .1  & 1&284  & MeerKAT   & 15    & 66 & .8  & $\pm$ & 6 & .7  \\
2021-09-26.63 & 59483 & .6      & 49 & .1  & &816   &  MeerKAT  & 15    & 72 & .0  & $\pm$ & 7 & .2 \\
2021-09-26.65 &  59483 & .6     & 49 & .1  & & \; 54    & LOFAR     & 120   & 61 & .4  & $\pm$ & 29 & .7  $^{*}$ \\
2021-09-26.73 & 59483 & .7      & 49 & .2  & &154   & LOFAR     & 120   & 49 & .8  & $\pm$ & 7 & .4 \\
2021-10-11.63 & 59498 & .6      & 64 & .1  & &816   & MeerKAT   & 15    & 59 & .2  & $\pm$ & 5 & .9 \\
2021-10-11.67 & 59498 & .7      & 64 & .2  & 1&284  & MeerKAT   & 15    & 53 & .6  & $\pm$ & 5 & .4 \\
2021-10-12.63 & 59499 & .6      & 65 & .1  & &154   & LOFAR     & 120   & 52 & .0  & $\pm$ & 6 & .5 \\
2022-03-17 $^{**}$ & 59655 &     & 220 & .5 & &\; 54    & LOFAR     & 360   & 13 & .1  & $\pm$ & 13 & .3 $^{***}$\\
2022-03-17.22 & 59655 & .2      & 220 & .7  & 1&284  & MeerKAT   & 15    & 11 & .6  & $\pm$ & 1 & .2 \\
2022-03-17.26 & 59655 & .3      & 220 & .8 &  &816  & MeerKAT   & 15    & 12 & .2  & $\pm$ & 1 & .2 \\
2022-03-19.19 & 59657 & .2      & 222 & .7 & &154   & LOFAR     & 120   & 12 & .7  & $\pm$ & 3 & .4 \\
2022-03-19.20 & 59657 & .2      & 222 & .7 & 1&284  & MeerKAT   & 15    &  11 & .5  & $\pm$ & 1 & .2 \\
2022-03-19.24 & 59657 & .2      & 222 & .7 & &816   & MeerKAT   & 15    & 11 & .9  & $\pm$ & 1 & .2 \\

\hline 

\end{tabular}
\begin{tablenotes}\footnotesize
\item[*] Note that this only a marginal detection at $3.5$ x rms noise.
\item[**] These $6$ hours were spread over three observation dates: 2022-03-16 04:47:31, 2022-03-17 04:53:35 and 2022-03-18 04:39:39.
\item [***] $13.07 ~\pm ~ 13.22$ represents the force extracted flux value. The $3\times\rm{rms}$ upper limit is $22.26$ mJy.
\end{tablenotes}
\end{threeparttable}
\end{table*}

\section{Results}
\label{sec:results}
\subsection{Radio light curves}

\begin{figure}
    \centering
    \includegraphics[width=\linewidth]{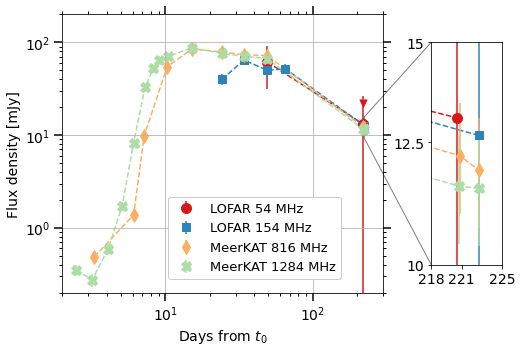}
    \caption{Radio light curves for the RS Oph 2021 eruption. Data presented here are given in Table \ref{tab:obs}. For the final LOFAR LBA observation both the $3\sigma$ upper limit (downwards pointing triangle) and the forced flux extraction (including large error bars) are given. $t_0$ is taken as MJD $59434.5$.}
    \label{fig:lightcurve}
\end{figure}

The flux density evolution of RS Oph during its $2021$ outburst, at observing frequencies of $54$, $154$, $816$ and $1284$ MHz, is shown in Figure \ref{fig:lightcurve}, based on the observations described in Section \ref{sec:observations} and summarised in Table \ref{tab:obs}. Radio emission was first detected from the nova on day $2.5$ at $1.28$ GHz. RS Oph is consistent with a point source in all epochs and frequencies, except for the LOFAR LBA observations on day $220$, which only resulted in an upper limit.

The MeerKAT light curves first show a brief flat flux density phase, followed by a steep rise in flux, a `plateauing' phase at maximum, where the flux density seems approximately constant over a period of $\sim 30$ days, followed by a steep decline. The first few data points of the $1.28$ GHz light curve show a stable (or slightly decreasing) flux density and then a sharp increase at day $4$, indicating that the non-thermal emission only became visible or dominant around this time. The slope in the rise of the flux density over the first $10$ days seems to break to an even steeper rise around day $5-6$. The LOFAR observations were obtained via a DDT proposal and therefore only begin at day $24$. Figure \ref{fig:lightcurve} is consistent with radio observations of other novae, where different observing frequencies peak at different times, with lower frequencies turning on at later times. The general trend of the LOFAR $154$ MHz light curve seems similar to the general trend observed with MeerKAT; `plateauing' and steep decline. The data points around day $220$ reveal a change in the slope of the decline after peak flux density, where the flux density seems to decline much steeper at late times. In Section \ref{sec:lc_modelling} we perform detailed modelling to explore the various emission components that contribute to this light curve. In Section \ref{sec:lc_modelling2} a fit is performed to determine the mass loss from the red giant wind.

\subsection{Spectral evolution}

\begin{figure*}
    \centering
    \includegraphics[width=\linewidth]{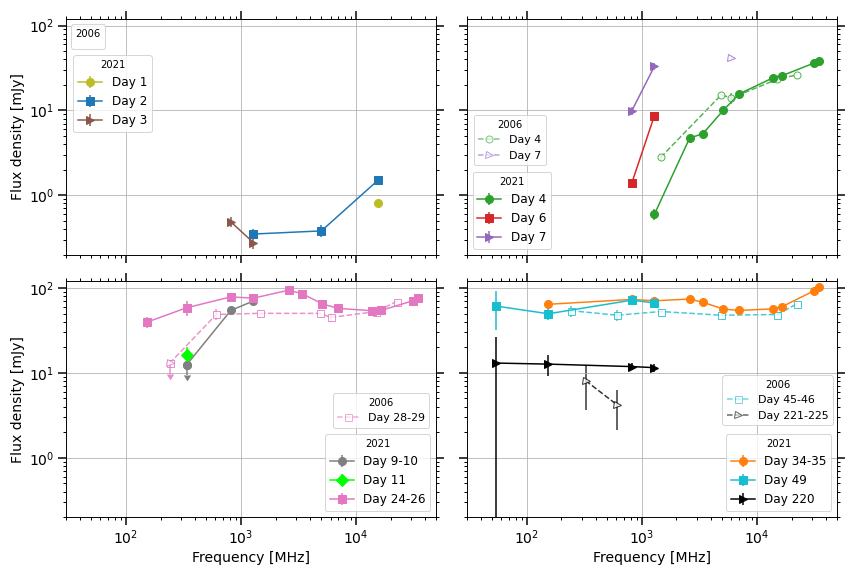}
    \caption{Spectrum on various days of the $2021$ outburst (solid lines) and the $2006$ outburst (dashed lines and open markers). The four panels represent four different epochs of the outburst, days $1$ to $3$, days $4$ to $7$, days $9$ to $29$, and days $34$ to $225$ respectively. LOFAR and MeerKAT data are presented in this paper (see Table \ref{tab:obs}), data from other facilities and the $2006$ outburst are listed in Appendix \ref{ap:additional_data}. }
    \label{fig:rsoph_spectrum}
\end{figure*}

Using a selection of the MeerKAT and LOFAR observations supplemented by observations by VLITE \citep{ATEL_vlite}, \textit{e}-MERLIN, AMI-LA \citep{ATEL_ami} and the VLA \citep{ATEL_vla}, we evaluate the evolution of the spectrum over time. The spectrum for different days after the outburst is shown in Fig. \ref{fig:rsoph_spectrum}. For reference, the spectral evolution for a selection of days during the $2006$ outburst is also shown \citep{kantharia2007giant, eyres2009double}. The data points from other facilities and the $2006$ outburst are summarized in Tables \ref{tab:obs_atels_2021} and \ref{tab:obs_2006} in Appendix \ref{ap:additional_data}.

The spectral evolution of the $2021$ outburst seems very similar to the $2006$ outburst, showing an extremely quick rise in the flux density at high frequency. Emission at low radio frequencies rises less quickly, but low-frequency observations with LOFAR only began around day $24$. A notable feature of the spectrum is the inverted spectrum at day $3$, where the spectral index for the MeerKAT observations is $\alpha = -1.25^{+0.67}_{-0.67}$ ($S_\nu \propto \nu^{\alpha}$), indicating that the radio emission is non-thermal. The negative spectral index observed on day $3$ changes to a positive spectral index on day $4$. This implies that a high-frequency, thermal emission component quickly overwhelms the low-frequency non-thermal component. This high-frequency component rises over the course of the next week and the absorption (turnover) frequency slowly moves towards lower frequency. After this early evolution, the spectrum stabilizes between day $24$ and $49$. At this time the spectrum is quite flat over two orders of magnitude in frequency, from $50$ MHz to $5$ GHz. The final measurement at day $220$ shows an extremely flat spectrum, where the flux density has decreased at all frequencies. This development is similar to the spectral progression of the $2006$ outbursts, where a low-frequency (non-thermal) emission component dominates below $1.4$ GHz from day $20$ \citep{kantharia2007giant, eyres2009double}. This can be seen from the negative spectral index below 1 GHz in the day 45-46 spectrum from the 2006 outburst. Other low-frequency data points of the 2006 outburst indicate a spectral index that evolves from $\alpha \approx -0.1$ to $\alpha \approx -1.0$ \citep{kantharia2007giant}. At higher frequency, a thermal component, originating from an expanding, decelerating shell of ejecta, dominates \citep{sokoloski2008uncovering}. 

The spectral shape has been interpreted as the result of several different emission components in previous outbursts \citep{kantharia2007giant, eyres2009double}. In Section \ref{sec:sed_modelling}, we will interpret the spectral evolution in this light. However, the extremely flat spectrum of RS Oph around day $20-50$ is also reminiscent of a jet spectrum \citep{blandford1979jets, markoff2001jet, moscibrodzka2013coupled}. Additionally, an extremely collimated thermal outflow has been observed in the $2006$ outburst \citep{sokoloski2008uncovering}. In Section \ref{sec:jet} we will apply the simple steady jet model, as described in \cite{blandford1979jets}, to the RS Oph radio observations to examine whether the observed radio emission could originate from a synchrotron jet.

\section{Discussion}
\label{sec:discussion}

\subsection{Brightness temperatures}
\label{subsec:brightness_temp}

The previous sections presented the lowest frequency observations of a (RS Oph) nova eruption to date. This allows us to place stringent constraints on the brightness temperature.

In the following, we assume a spherical emitting volume and define the brightness temperature as:

\begin{equation}
    T_b = \frac{S_{\nu} c^2}{2k_B\nu^2}\frac{d^2}{\pi R^2}
    \label{eq:brightness_temp}
\end{equation}
with $d$ being the distance to RS Oph and $R$ the radius of the emitting region. The distance to RS Oph is subject of debate \citep{barry2008distance, schaefer2009orbital} with values ranging from $\sim 1$ kpc to $>5$ kpc. Recently however, Gaia DR3 parallax measurements settle that debate and place RS Oph at a distance to be $2.68 ^{+0.17}_{-0.15}$ kpc (see also \citealt{schaefer2022comprehensive}). This distance also agrees very will with the distance derived in \cite{rupen2008expanding} using expansion parallax techniques in VLBA images. In the remainder of the discussion, we use this distance and provide distance-scaled formulae to allow for a different distance assumption. We note that the assumption of a single expansion velocity and spherical emitting volume is not correct, since we know from VLBI imaging that the source is not spherically symmetric \citep{munari2022radio}. However, we choose to still use this approximation as it is standard procedure when estimating the brightness temperature of an unresolved source.

The size of the emission region at a given time can be calculated with the (average) expansion velocity of the ejecta. \cite{munari2022radio} show that the average expansion velocity of the ejecta is $7500 \pm 150 \; \rm{km}~\rm{s}^{-1}$. This was found by analyzing the angular extent of the ejecta. This compares well to the values found by HST imaging of the $2006$ ejecta presented by \cite{ribeiro2009expanding}. Estimates presented from spectral line observations (see eg.  \cite{ATEL_vel_spec, taguchi2021atel,ATEL_vel_spec3,ATEL_vel_spec4,  fajrin2021atel}) measure just the radial component of the expansion velocity, while the \cite{munari20212021} estimate captures both the radial and tangential component averaged over the first 34 days of the outburst.

\begin{figure}
    \centering
    \includegraphics[width=\linewidth]{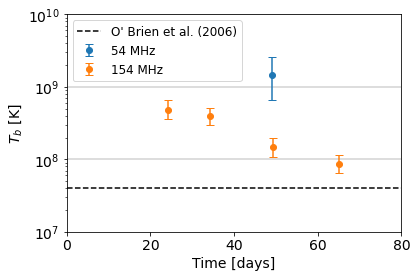}
    \caption{Brightness temperatures of the RS Oph $2021$ outburst corresponding to LOFAR observations at different times after the outburst. Here a distance of $d =2.68 ^{+0.17}_{-0.15}$ kpc and an expansion velocity of $7500 \pm 150$ km s$^{-1}$ are assumed. The error bars account for uncertainties in observed flux density, distance and expansion velocity. The dashed black line shows the previous best estimates of the brightness temperature \citep{o2006asymmetric}.}
    \label{fig:brightness_temp}
\end{figure}

Figure \ref{fig:brightness_temp} shows the brightness temperature calculated for the LOFAR observations assuming an expansion velocity of $v_{\rm exp} \sim 7500~{\rm km~s^{-1}}$ and a distance to RS Oph of $2.68$ kpc. We only present the LOFAR observations in this figure, since they are at lower frequency than MeerKAT and therefore place the strongest constraints on the brightness temperature.

The LOFAR LBA observation on day $49$ gives the most constraining value on the brightness temperature. A flux density of $61.4 \pm 29.7$ mJy was observed at a frequency of $54$ MHz. This measurement implies a brightness temperature of $1.5 \cdot 10^9$ K. The error bars in Figure \ref{fig:brightness_temp} include the aforementioned errors on the distance, expansion velocity and flux density. For the LBA observation, this error interval is $T_{\rm b} = 1.5 ^{+1.1}_{-0.8} \cdot 10^9$ K. Using the LOFAR HBA observation on day $24$ yields a similar brightness temperature. A flux density value of $39.9 \pm 5.65$ mJy was measured at $154$ MHz $24$ days after outburst, implying $T_{\rm b} =4.8 \cdot 10^8 \; \rm{K}$.

The brightness temperature of $T_{\rm b} = 1.5 ^{+1.1}_{-0.8}\cdot 10^9$ K is over an order of magnitude higher than the values calculated from the previous outburst ($T_{\rm b} \approx 4 \cdot 10^7 $ K; \citealt{o2006asymmetric}). 
Note that the brightness temperature estimate by \cite{o2006asymmetric} is based on observations at 5 GHz, possibly diluting the measurement with thermal emission. The high brightness temperature at 154 MHz confirms that a non-thermal emission component dominates at low frequency during the nova explosion. Furthermore, this brightness temperature implies that there are at least mildly relativistic electrons present in the emitting plasma, since $kT=m_e c^2$ at $T= 5.9 \cdot 10^9$ K.

\subsection{Modelling of the radio light curves}
\label{sec:lc_modelling}

In this section, we will explore the interpretation of the $2006$ and $2021$ radio light curves of RS Oph. We also compare the light curves of the $2021$ outburst to the most recent outburst of V3890 Sgr, another recurrent nova with a red giant companion.

\begin{figure*}
    \centering
    \includegraphics[width=\textwidth]{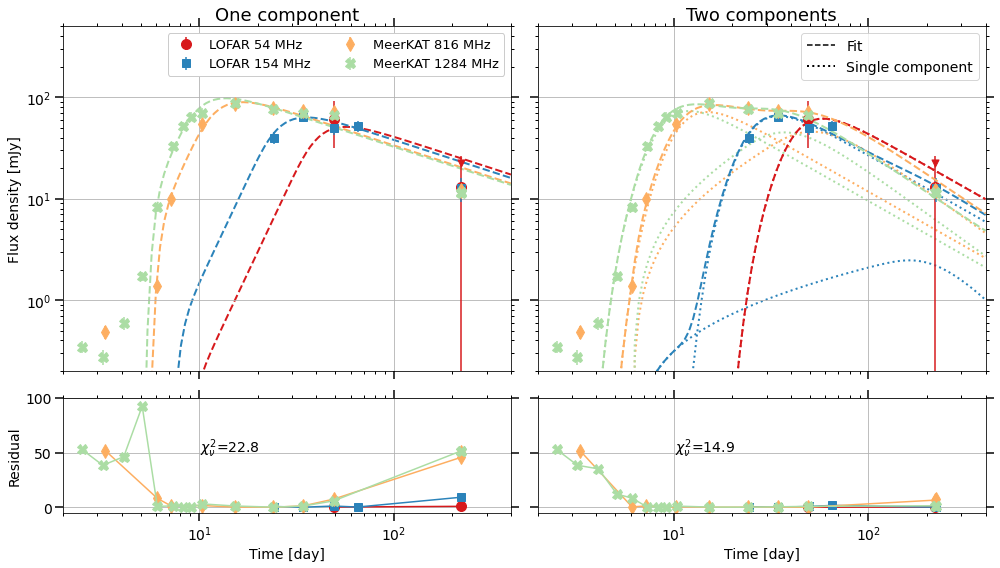}
    \caption{Radio light curves of the RS Oph $2021$ eruption including model fits. The left panel shows the best fitting radio supernova models, as described in Appendix \ref{ap:lightcurve_model}, as a dashed line. The right panel shows the fit resulting from a model where two synchrotron components are allowed. The dashed lines show the total fit, which is the sum of the two components shown with the dotted lines. The bottom panels show the residual, which is defined as $\left(O_i-C_i \right)^2 / \sigma_i^2$ where $O_i$ is the observed value, $C_i$ is the model value, and $\sigma_i$ is the error on the observed value. The reduced chi-squared statistic is described in Section \ref{sec:lc_modelling}. }
    \label{fig:light_curve_fits}
\end{figure*}

At first glance, the features of the light curves in Figure \ref{fig:lightcurve} are very similar to those of radio supernovae, and to the $2006$ outburst \citep{kantharia2007giant} where the observed light curves are  modelled with radio supernovae models developed by \cite{weiler1996radio, weiler2002radio}. Appendix \ref{ap:lightcurve_model} provides a detailed description of these models. In this scenario, the nova ejecta shock the circumbinary red giant material, which is ionized and heated by the nova, generating the relativistic electrons and enhanced magnetic fields, necessary for synchrotron emission \citep{chevalier1982radio, chevalier1982young}. The fast increase in radio flux is due to the decrease in the opacity of the circumstellar material in front of the shock. As the emission region expands, the optical depth from the red giant wind ahead of the shocked material decreases and the velocity of the shock front decelerates, which leads to fading radio emission. The left panel of Figure \ref{fig:light_curve_fits} shows the best-fitting radio supernova model fit to the light curves. The model is detailed in Appendix \ref{ap:lightcurve_model} and the explored parameter space and fit parameters are presented in Table \ref{tab:fit_values_one_component} in Appendix \ref{ap:fit_parameters}. The bottom panel shows the residual which is defined as $\left(O_i-C_i \right)^2 / \sigma_i^2$ where $O_i$ is the observed value, $C_i$ is the model value, and $\sigma_i$ is the error on the observed value.

It is clear that the simple radio supernova model does not fully capture the data. The early measurements ($t<5$ days) are not well-explained, just as the steep decline in flux we observe in our data at late times. In the next two sections, we will provide a possible explanation for the model's shortcomings at early and late times.

\subsubsection{Capturing the steep decline: a two-component synchrotron model}

Studying the light curves, one can see a 
plateauing of the radio fluxes around $t\sim 20-30$ days. The plateauing of the light curve (or increase in the case of the LOFAR $154$ MHz data) could be explained by a second synchrotron emitting component, as was found in the previous outburst (see section \ref{sec:2006}). Furthermore, multiple ejecta components are also suggested by the gamma-ray observations \citep{diesing2022evidence}. To test this hypothesis against our data, we fit the data again, allowing for two components of the radio supernovae  model \citep{weiler1996radio, weiler2002radio}. We note that there is no reason in particular that this second emission component should behave exactly similarly to the first emission component. However, using this approach we can test whether a two-component model would fit the data better. 
In the right panel of Figure \ref{fig:light_curve_fits} we show a fit of this two-component model to our data. The dashed line shows the full two-component model, while the dotted lines indicate the course of the individual components. Note that the model suggests that the second component is too faint at 54 MHz to be visible in this plot. The residuals in the bottom of this figure are defined as the absolute value of the difference between the model and the data ($\left(O_i-C_i \right)^2 / \sigma_i^2$) as explained in the previous section. Comparing these absolute residuals between the one and two-component fit shows that the two-component model captures the steep decline of the light curve at late times much better than the single-component model. The explored parameter space and fit parameters are presented in Table \ref{tab:fit_values_two_component} in Appendix \ref{ap:fit_parameters}.

We use the reduced chi-square statistic to test if the two-component model (right panel in Figure \ref{fig:light_curve_fits}) is preferred compared to the one-component model (left panel in Figure \ref{fig:light_curve_fits}). The reduced chi-square statistic is defined as
$\chi_{\nu}^2  = \frac{1}{\rm{DOF}} \sum_i \frac{\left(O_i-C_i \right)^2}{\sigma_i^2}$
where DOF is the degrees of freedom, defined as the number of observations minus the number of fitted parameters, $O_i$ is the observed value, $C_i$ is the model value, and $\sigma_i$ is the error on the observed value. The result of an F-test on these reduced chi-squared values shows that the data prefer the two-component model with an extremely small p-value of $4 \cdot 10^{-6}$. In conclusion, we are for the first time able to show that the low-frequency radio light curves prefer a delayed onset second emission component, emerging around day $20$ after the outburst. The onset of this second component in the light curve coincides with the time around which a second component was seen to separate from the central component with VLBI observations in $2006$ \citep{o2006asymmetric, rupen2008expanding}.

\subsubsection{Capturing the $t<5$-days data points: hints of a residual synchrotron component \label{sec:t<5_old_pop}}
We now investigate these models at early times. There are $15$ years between the previous and most recent outburst of RS Oph; extrapolating the current light curve model (right panel of Figure \ref{fig:light_curve_fits}) to $t=15$ years reveals whether or not the radio emission has completely decayed in between outbursts. The results of this exercise are presented in Table \ref{tab:extrapolate_flux}. It is clear that based on the models presented here, we do not expect the radio emission to have fully disappeared after $15$ years. This implies that the early flux measurements of the $2021$ outburst before the light curve starts rising, could be due to an old emission component from the $2006$ outburst. An old synchrotron component would be able to explain the inverted spectral index that was observed on day $3$, where the spectral index for the MeerKAT observations is $\alpha = -1.25^{+0.67}_{-0.67}$. A straightforward way to confirm this hypothesis would be to monitor RS Oph in the radio over the coming years. In the recurrent nova T CrB increased radio emission was also detected during optical quiescence, however in that case the optical emission also slightly increased by two magnitudes, pointing to a phase of higher accretion \citep{linford2019t}. 

We note that this extrapolation assumes a uniform circumbinary medium, which might not be the case if the red giant wind material has an edge. Jumps in the density could possibly be introduced by previous nova outbursts. A decreased density by successive outburst has been suggested \citep{kantharia2015insights}, but to our knowledge, no observations of the detailed large-scale circumbinary density profile have been made. Theoretical work by \cite{moore2012circumstellar} on the circumbinary medium suggests that the ejecta shell should sweep up the entire wind before the next outburst. However, predictions in this work are in tension with our late-time detection of the system on day 220. Monitoring the decay of the light curve and possibly observing a break when the ejecta reach a density differential could (via the formalism introduced by \cite{moore2012circumstellar}) help put further constraints on the ejecta mass and red giant mass loss rate.

\begin{table}
\begin{tabular}{l|ll}
            & Flux (mJy)    & Flux (mJy) \\ 
            & $t=15$ years  & $t<5$ day (2021) \\  \hline
LOFAR 54 MHz        & 0.69      & -                      \\
LOFAR 154 MHz       & 0.46      & -                      \\
MeerKAT 816 MHz     & 0.28      & $0.486 \pm 0.068 $     \\
MeerKAT 1284 MHz    & 0.22      & $0.349 \pm 0.048 $                   
\end{tabular}
\caption{The flux at t=$15$ years, calculated by extrapolating the two-component synchrotron models in Fig. \ref{fig:light_curve_fits}. The third column shows the flux density as measured before the sharp rise of the light curve at $t<5$ days.}
\label{tab:extrapolate_flux}
\end{table}

\subsubsection{Origin of second component / double-peaked light curves}
\label{sec:double_peak}

\begin{figure}
    \centering
    \includegraphics[width=\linewidth]{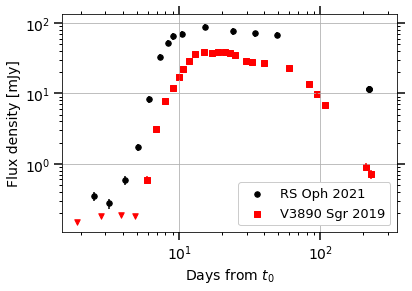}
    \caption{MeerKAT $1.28$ GHz observations of RS Oph in $2021$ and V3890 Sgr in $2019$ \citep{nyamai2021radio}. The red downwards pointing triangles indicate upper limits instead of detections for the V3890 Sgr outburst. }
    \label{fig:rs_oph_v3890_sgr}
\end{figure}

In the previous sections, we have presented a model that captures the plateauing (or double-peaked) behaviour of the light curve. Such double-peaked behaviour has been observed in recurrent novae in the past. \cite{eyres2009double} show $1.46$ GHz light curves obtained with the VLA from the $2006$ RS Oph outburst. This light curve shows a double-peaked behaviour, which the authors suggest could be explained by a second emission component. In Section \ref{sec:sed_modelling}, we discuss the various emission components that could contribute to the observed radio emission in depth. However, in this section, we compare the RS Oph light curve with the double-peaked light curve that was observed in the most recent outburst of recurrent nova V3890 Sgr \citep{nyamai2021radio}. We compare the light curve of V3890 Sgr to RS Oph since both have red giant companions and both are long period binaries, with periods of around $520$ and $454$ days respectively \citep{schaefer2010comprehensive}. Figure \ref{fig:rs_oph_v3890_sgr} shows the remarkable similarity between the MeerKAT $1.28$ GHz light curve of the $2021$ RS Oph outburst in the black dots and the $2019$ V3890 Sgr outburst in red squares. 

In V3890 Sgr, the spectral index indicates that the second radio bump is dominated by non-thermal emission \citep{nyamai2021radio}. If we assume that the second radio bump is from the original shock wave, the plateauing of the radio flux could be explained by either an increase in the velocity of the shock wave or an increase of the density of the material being shocked. 3D simulations of the outburst of RS Oph indicate that the surrounding medium is not uniformly distributed \citep{booth2016modelling}. Hence an increase in the density of the surrounding medium is possible and could give rise to plateauing light curves. Additionally, the surrounding medium might be asymmetric. For example, the nova ejecta could first hit an equatorial and then a polar density jump (see eg. \cite{chomiuk2014binary}), which would explain the double-peaked behaviour in the light curve. Detailed modelling of the density distribution in the surrounding medium is not feasible at this point. In the next section, we will make simplifying assumptions to estimate the mass-loss rate of the companion star.

\subsection{Estimating mass-loss rate from the red giant companion}
\label{sec:lc_modelling2}

Using a simple circumstellar material (CSM) profile, we can estimate the mass-loss rate from the red giant companion. The radio light curve of RS Oph is compared to non-thermal emission produced through the blast-wave mechanism of supernovae \citep{chevalier1998synchrotron}. Details of how the model is applied to explain radio emission from novae are presented in \citet{nyamai2021radio}. The model depends on the mass of the ejected envelope ($M_{\rm ej}$), the radial ($R_{\rm shock}$) and velocity ($V_{\rm shock}$) profile of the ejecta, velocity of the red giant wind ($V_{\rm wind}$) and spectral index of electrons producing non-thermal emission ($p$).

Different studies following the nova outbursts in $1985$, $2006$ and $2021$ estimate an ejecta shell mass of $3-5~\cdot 10^{-6}~\textrm{M}_\odot$ \citep{hjellming1986radio, das2015abundance, pandey2022study}. Here, we consider an ejecta of mass $M_{\rm ej} = 4~\cdot 10^{-6}~\rm{M}_\odot$. The ejecta mass is not varied in the modelling since it is relatively well-constrained. Some tests were performed to vary the ejecta mass within the range $M_{\rm ej} = 3-5~\cdot 10^{-6}~\rm{M}_\odot$ for a given mass-loss rate, but this does not make any significant changes for the predicted light curves. Given that the compact object orbits within the red giant wind, the radius and velocity of the radio luminosity producing blast wave depend on the density of the nova ejecta and the density of the circumbinary material. The dynamics of such a shockwave have been described in \citet{nyamai2021radio}. The presence of a shock in RS Oph has been detected during the $1985$, $2006$ and $2021$ outbursts \citep{hjellming1986radio, sokoloski2006x, pandey2022study}. The evolution of the shock wave is described by two phases, the free expansion where the ejected mass is much larger than the surrounding medium and the Sedov-Taylor phase where the swept-up material increases significantly \citep{bode1985model, Tang2017}. Theoretically, during the free expansion phase, $R_{\rm shock} \propto t$ and $V_{\rm shock}$ are constant. During the Sedov-Taylor phase, $R_{\rm shock}\propto t^{0.67}$ and $V_{\rm shock}\propto t^{-0.33}$ \citep{nyamai2021radio}. These estimates depend on the density profile of the ejecta as $\rho_{\rm ej} \propto r^{-2}$, where $r$ is the radial distance from the compact object.

For the assumed ejecta mass, the kinetic energy of the nova ejecta is determined by the initial velocity of $4700~\rm{km~s^{-1}}$ on day $2$ as $2.5~\cdot 10^{44}~\rm{erg}$. The velocity of the ejecta was determined using H$\alpha$ emission line profiles \citep{munari2022radio}.  Furthermore, equipartition energy in accelerated particles and amplified magnetic fields is assumed such that $\epsilon_e$ = $\epsilon_B$ = $0.001$ is assumed. Lowering the values of the microphysical parameters would increase the observed flux density but these values are already on the low end (for more details see \cite{nyamai2021radio}) and therefore we choose not to vary them here. The mass-loss rate of the red-giant wind is estimated using the fit of the radio luminosity during the rise of the radio light curve.

\begin{figure}
    \centering
    \includegraphics[width=\linewidth]{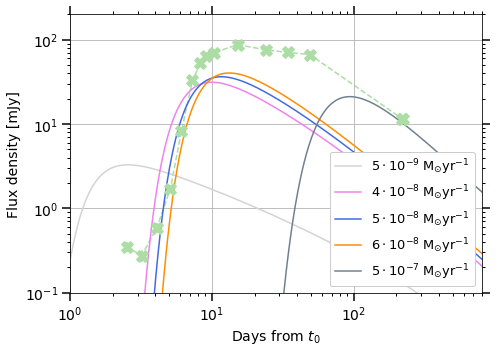}
    \caption{MeerKAT $1.28$ GHz observations of RS Oph following the $2021$ outburst (green crosses) and a model of radio emission due to non-thermal emission undergoing free-free absorption for various red giant mass-loss rates.}
    \label{fig:rs_oph_modelling}
\end{figure}

Using the input parameters:$p=2.1$, $V_{\rm wind} = 20~{\rm km~s^{-1}}$ \citep{walder20083d}, $d=2.68~{\rm kpc}$, we test if the determined mass-loss rate produces synchrotron emission that matches the radio light curve. \cite{booth2016modelling} estimated a mass-loss rate of $5 \cdot 10^{-7} \; {\rm M_\odot} {\rm yr}^{-1}$ in order to produce an ionization structure that is consistent with observations. A similar wind mass loss rate is predicted by \cite{diesing2022evidence}. We therefore first test a mass-loss rate of $5 \cdot 10^{-7}~{\rm M_\odot~yr^{-1}}$ and find that the model produces a light curve that appears at a much later time than observed. Subsequently, we test $5 \cdot 10^{-8}~{\rm M_\odot~yr^{-1}}$, and $5 \cdot 10^{-9}~{\rm M_\odot~yr^{-1}}$. Figure \ref{fig:rs_oph_modelling} shows the predicted light curves for the different red giant mass-loss rates. The three mass-loss rates produce wildly different light curves and we find that a mass-loss rate of $5 \cdot 10^{-8}~{\rm M_\odot~yr^{-1}}$ produces the closest match to the observations in terms of when the rise of the emission occurs. We refine our search by testing mass-loss rates of  $4 \cdot 10^{-8}~{\rm M_\odot~yr^{-1}}$, $5 \cdot 10^{-8}~{\rm M_\odot~yr^{-1}}$ and $6 \cdot 10^{-8}~{\rm M_\odot~yr^{-1}}$, the results of which are shown in Figure \ref{fig:rs_oph_modelling}. We decide to not test our models with any goodness-of-fit statistic as Figure \ref{fig:rs_oph_modelling} clearly shows that none of the models fit the data well. The $5 \cdot 10^{-8}~{\rm M_\odot~yr^{-1}}$ model agrees with the data during the rise of the radio light curve but does not predict the early emission, < $4$ day and the peak, and the decay phase of the light curve (see Figure \ref{fig:rs_oph_modelling}).

The light curve models presented in this section favour a mass-loss of around $5 \cdot 10^{-8}~{\rm M_\odot~yr^{-1}}$ for the MeerKAT data, which an order of magnitude lower than previous mass-loss rate estimates of around  $5 \cdot 10^{-7}~{\rm M_\odot~yr^{-1}}$ \citep{booth2016modelling, diesing2022evidence}. However, from Figure \ref{fig:rs_oph_modelling} it is also clear that none of the simulated light curves represent the data well.
It is not surprising that three phases of the radio light curve cannot be described by a simple model since we assume a uniform distribution of the ejected material and radio imaging shows non-spherical nova ejecta \citep{munari2022radio}. Possible sources of uncertainty include the presence of thermal emission in addition to the synchrotron emission. Additionally, the microphysical parameters $\epsilon_e$ and $\epsilon_B$ are poorly constrained and understood \citep{nyamai2021radio}, and a simple assumption of $\epsilon_e = \epsilon_B = 0.001$ is made here. Future modelling should consider emission distinctively from the two emission components and any residual emission from the previous outbursts.

\subsection{Emission components and spectral modelling}
\label{sec:sed_modelling}

\begin{figure}
    \centering
    \includegraphics[width=\linewidth]{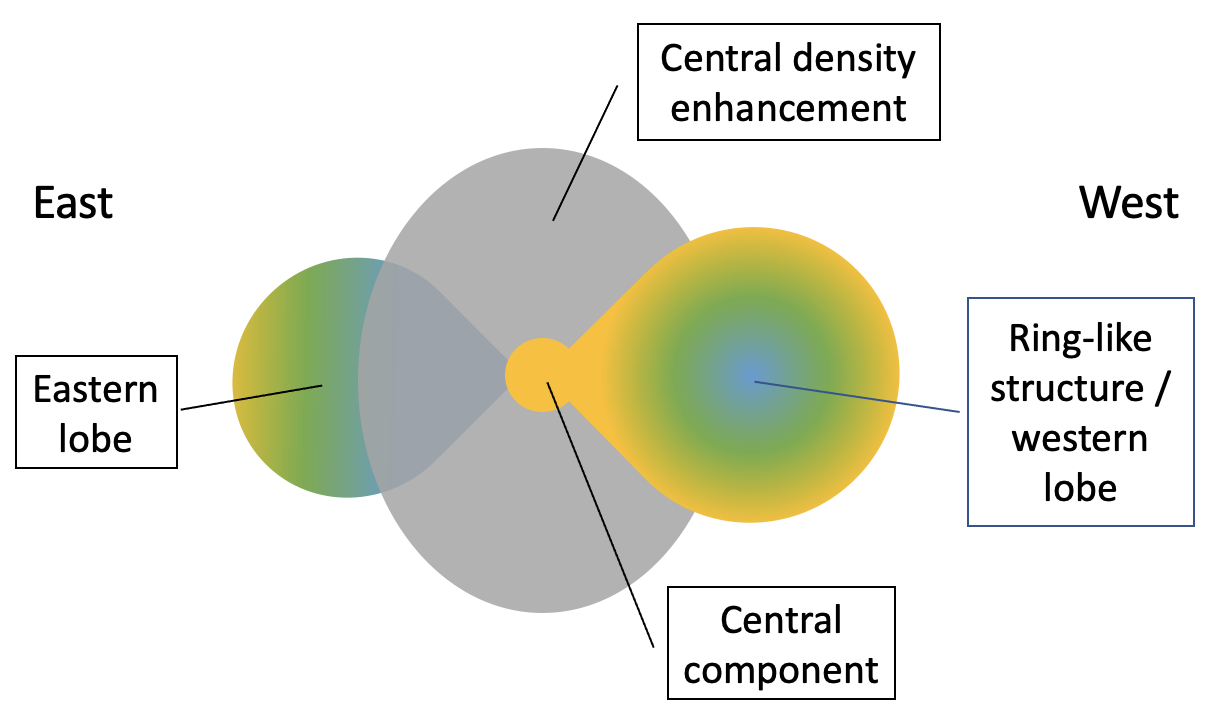}
    \caption{Simple geometry sketch of RS Oph (not to scale) of the expanding and bipolar arrangement of the RS Oph ejecta. The radio brightest parts of the eastern and western lobe are shown in yellow. This sketch is largely based on \protect\cite{munari2022radio}, but the different components are also named in accordance with work by \protect\cite{o2006asymmetric, rupen2008expanding, sokoloski2008uncovering}. }
    \label{fig:geometry_sketch}
\end{figure}

In this section, we discuss the emission components that could contribute to the radio emission. Using the MeerKAT and LOFAR observations presented in this work, supplemented by observations by VLITE \citep{ATEL_vlite}, \textit{e}-MERLIN, AMI-LA \citep{ATEL_ami} and the VLA \citep{ATEL_vla}, we evaluate the evolution of the spectrum over time, as presented in Figure \ref{fig:rsoph_spectrum}. First, we will summarize the spectral modelling work from the $2006$ outburst. To guide the reader a simple geometry sketch is included in Figure \ref{fig:geometry_sketch}, showing the radio brightest components in yellow. This Figure is largely based on the EVN observations by \cite{munari2022radio}, but different components are also named in accordance with work on the 2006 outburst.

\subsubsection{$2006$ outburst}
\label{sec:2006}

Detailed analysis of the previous outburst of RS Oph shows that there are multiple radio-emitting components. \cite{o2006asymmetric} show the outburst results in an expanding shock wave as it sweeps through the red giant wind, producing a remnant similar to that of type II supernova, but evolving over a much shorter timescale. The radio emission is non-thermal synchrotron emission and the remnant is rather non-spherical. Furthermore, a second component seems to emerge to the east of the ring. This same two-component structure is also found in \cite{rupen2008expanding}. They describe an expanding ring, most naturally associated with the shock wave resulting from the outburst and an additional synchrotron emitting component well to the east of the shell, see Figure \ref{fig:geometry_sketch}. The central part of the ring-like structure is the brightest and has a spectral index $\alpha \approx -0.3$ at day $27$. The west (far) side of the ring has a spectral index $\alpha \approx -0.7$, similar to the additional synchrotron emitting component to the east of the ring with a spectral index $\alpha \approx -0.7$ at day $27$. The west side of the ring and the second component to the east are well explained by an optically thin synchrotron spectrum, because spectral indices generally range from $\alpha \approx -0.6$ to $\alpha \approx -1.0$ for optically thin synchrotron emission. Despite the quite flat spectral index, the central part of the shell-like component is best explained by optically thin synchrotron emission. Thermal bremsstrahlung is ruled out as it would require extremely large thermal energies. It is likely that emission from the central part is a mix of thermal and non-thermal emission processes.
Finally, \cite{sokoloski2008uncovering} find the same aforementioned two components, but additionally observe a highly collimated thermal outflow (jet) at $43$ GHz. The half-opening angle is determined to be less than approximately $4$ degrees. The specific intensity of the jet is consistent with thermal bremsstrahlung (free-free) emission, from a gas with a temperature of T>$10^4$ K. They also interpret the second component to the east as a synchrotron lobe and they determine the spectral index to be $-0.53$. 

In summary, there are likely to be at least four emission components. The central, brightest radio component is likely due to non-thermal synchrotron emission, but the unusual spectral index suggests that a thermal component also contributes to the radio emission at this location. The (west) far side of the ring is consistent with a synchrotron emitting component. The blob to the east of the central component emerges around day $20$ and the spectral index shows that this component is also consistent with synchrotron emission. At higher frequencies, collimated outflows are observed that are clearly thermal \citep{sokoloski2008uncovering}.

\subsubsection{$2021$ outburst}

The spectrum presented in Figure \ref{fig:rsoph_spectrum} shows that the evolution of the spectrum of the $2021$ outburst is similar to the $2006$ outburst. Additionally, \cite{munari2022radio} present high-resolution long-baseline EVN observations that show a very similar structure to the $2006$ outburst. An elongated structure over the east-west direction with a total extension of about $90$ mas is observed. A central compact component lies at the location of the Gaia DR3 position of the binary location, see Figure \ref{fig:geometry_sketch}. The radio emission is brighter on the western side, forming a circular lobe, while fainter emission is present to the east of the central component \citep{munari2022radio}. The authors explain this structure by a combined effect of an accretion disc and a density enhancement on the orbital plane that confine the nova ejecta primarily within a bipolar structure, expanding perpendicular to the orbital plane. The western lobe moves towards us, in the foreground to the central density enhancement, while the eastern lobe is expanding behind this density enhancement and therefore only emerges later in the outburst. 

In light of this interpretation, it is difficult to model the SED, as it is likely a combination of at least four emission components. Thermal and non-thermal emission at the location of the central object, a non-thermal component to the west that is visible from the start of the outburst, and finally, a non-thermal component to the east that is first obscured by the central density enhancement. The higher frequency (> $1$ GHz) emission at early times is likely a combination of thermal and non-thermal emission with an absorption frequency that evolves with time as the nova ejecta move outwards to a lower-density medium. 

At early times the MeerKAT spectrum at day $3$ points to a non-thermal component. Assuming that this non-thermal component rises at the lowest frequencies over the outburst, early time monitoring at low frequencies would possibly allow one to constrain the synchrotron self-absorption frequency. Finally, as suggested in Section \ref{sec:t<5_old_pop}, additional low-frequency monitoring at very late times could help monitor whether this synchrotron component actually disappears between outbursts. There has not been a radio monitoring campaign during quiescence in the past, but such an effort might reveal that there is a long-lasting old non-thermal emission component. These old synchrotron lobes might not actually disappear between outbursts.

\subsection{Equipartition magnetic field strength}
\label{sec:equipartition}

\begin{figure}
    \centering
    \includegraphics[width=\linewidth]{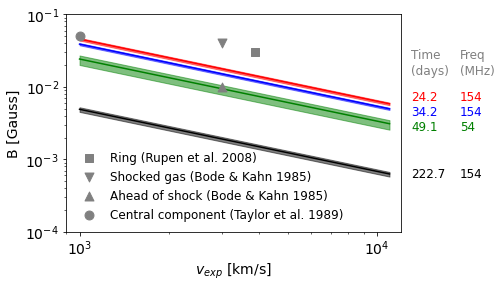}
    \caption{Lower limit on the equipartition magnetic field as a function of expansion velocity for various LOFAR flux measurements (see Table \ref{tab:obs} for details). Assuming the flux is largely comprised of optically thin synchrotron emission, equipartition between particles and the magnetic field and an electron/ion ratio $\eta=1$. The errorbands show uncertainty in flux.}
    \label{fig:b_field}
\end{figure}

Due to the large errors on the LOFAR LBA measurement at day 49, it is difficult to determine whether the spectrum of RS Oph is flat over the full frequency range or whether we observe  synchrotron self-absorption at the low end of the SED, see Fig. \ref{fig:rsoph_spectrum}. However, we can assume equipartition between the energy in electrons and energy in the magnetic field to estimate the magnetic field strength. Here we assume that the flux measured at LOFAR frequencies is (at least partly) comprised of an optically thin synchrotron emitting component. In the following, we assume that the main energy holding components are the electrons and the magnetic field and that the emitting volume is equal to the total volume of the astrophysical source (ie. a filling factor of one). Following \cite{longair2010high}, we calculate the magnetic field at minimum energy (the equipartition magnetic field) as 
\begin{equation}
B > 9\cdot10^3 \left(\frac{\eta L_{\nu}}{V} \right)^{2/7} \nu^{1/7}
\label{eqn:bfield}
\end{equation}
in cgs units; $\eta$ is the ion/electron ratio, $L_{\nu}$ in $\rm{erg}~ \rm{s}^{-1} \rm{Hz}^{-1}$, $V$ is the volume in $\rm{cm}^{-3}$, $\nu$ the frequency in Hz, and $B$ is the equipartition magnetic field in Gauss. Here we can replace the luminosity by $L_{\nu} = 4\pi D^2 F_{\nu}$ and $V=\frac{4}{3} \pi R^3$ with $D$ the distance to the source, $F_{\nu}$ the flux measurement in $\rm{erg} \rm{s}^{-1} \rm{Hz}^{-1} \rm{cm}^{-2}$ and $R$ the radius of the emission region in cm. As in Section \ref{subsec:brightness_temp}, we note that the assumed spherical symmetry is an approximation, as VLBI measurements show bipolar synchrotron blobs \citep{munari2022radio}. However, for the purpose of estimating the lower limit in the magnetic field we opt to use this approximation of spherical symmetry as this will likely overestimate the volume of the emission region, and therefore underestimate the magnetic field, which is appropriate as Eqn. \ref{eqn:bfield} calculates a lower limit on the magnetic field regardless. Furthermore, the size of the emitting region might decrease by a factor of a few, but this effect is suppressed by the power $(2/7)$ and does therefore not significantly impact the resulting magnetic field. Finally, calculating the size of the emission region by $R=v_{\rm{exp}} \cdot t$ and converting to more useful units, we find:

\begin{equation}
\begin{split}
    & B > 8.5 \cdot 10^{-3} \; \eta^{2/7} \left[\frac{F_{\nu}}{61.4 \; \rm{mJy}} \right]^{2/7} \left[\frac{D}{2.68 \; \rm{kpc}} \right]^{4/7} \\
    & \times{}  \left[\frac{v_{\rm{exp}}}{5000 \; \rm{km/s}} \right]^{-6/7} \left[\frac{t}{49.1 \; \rm{days}} \right]^{-6/7} \left[\frac{\nu}{54 \; \rm{MHz}} \right]^{1/7} \; \rm{G}
\end{split}
\end{equation}

with $\eta$ is the ion/electron ratio, $F_{\nu}$ the source flux, $D$ the distance, $v_{\rm{exp}}$ the expansion velocity, t the time since the nova eruption and $\nu$ the frequency. Ignoring the protons entirely, setting $\eta = 1$, allows us to calculate the lower limit on the equipartition magnetic field strength as a function of expansion velocity for various low-frequency flux measurements, as is shown in Figure \ref{fig:b_field}. The grey markers indicate previous estimates of the magnetic field by \cite{rupen2008expanding}, \cite{taylor1989vlbi} and \cite{bode1985model}. Note that for clarity we do not include the day 61 LOFAR observation in this plot. The equipartition magnetic field derived in \cite{rupen2008expanding} is placed at the expansion velocity corresponding to the velocity of the ejecta as measured by the VLBA in their work.

Figure \ref{fig:b_field} shows that observations around the peak of the light curves indicate a magnetic field strength between 5 and 50 mG.  These measurements agree with previous estimates of 10-50 mG \citep{bode1985model, taylor1989vlbi, rupen2008expanding} at moderate expansion velocities. Noting the work by \cite{beck2005revised}, where a revised equipartition formula for the magnetic field strength is presented, we find that this would give a revised magnetic field strength estimate of about 0.5 times smaller. This correction is much smaller than the uncertainty in the expansion velocity, and therefore not taken into account.

\section{Alternative scenarios for producing radio emission} \label{sec:alternative_scenarios}

In the previous section, we show that the radio light curve can be modelled by the shock interaction of the nova ejecta with the circumbinary red giant material. Furthermore, a non-uniform circumbinary medium or additional emission components can explain the plateauing of the light curve. The spectrum is likely a complex composition of various emission components. In this section, we show that stellar winds and a radio synchrotron jet can be confidently ruled out as origins for the radio emission.

\subsection{Stellar winds}

\begin{figure}
    \centering
    \includegraphics[width=\linewidth]{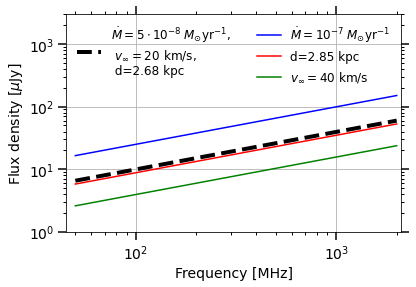}
    \caption{Radio flux from the red giant stellar wind. Black dashed line shows the wind parameters as estimated in Section \ref{sec:lc_modelling2} and by \protect \cite{walder20083d}. Coloured lines show how changing one parameter influences the radio flux.}
    \label{fig:stellar_wind}
\end{figure}

Stellar winds can contribute to a significant fraction of the observed radio emission for both single stars and stars in binary systems. In this section, we examine what fraction of radio emission is due to the stellar wind of the red giant companion in RS Oph. The radio emission from winds is either thermal bremsstrahlung emission from the ionized gas in the wind or non-thermal emission from shocks at the edges of the wind. We note that generally the wind of a red giant star is not expected to be ionized, however, for RS Oph it is suggested that the red giant wind gets (at least partially) ionized by the initial UV flash from the thermonuclear explosion and the radiation from the shock produced by the ejecta passing through the wind. For example \cite{rupen2008expanding} have assumed a fully ionized spherically symmetric wind and their modelling is in agreement with \cite{munari20212021}. The contribution to the radio flux from thermal wind emission can be estimated using the formalism derived by \cite{wright1975radio}:
\begin{equation}
\begin{split}
    S_{\nu} = & 200 \left(\frac{\nu}{\rm{5.5 \; GHz}} \right)^{0.6} \left(\frac{T_e}{10^4 \; \rm{K}} \right)^{0.1} \left(\frac{\dot{M}}{10^{-6} \; M_{\odot} \rm{yr}^{-1}} \right)^{4/3} \\
    & \times{} \left(\frac{\mu_e v_{\infty}}{\rm{100 \; km \; s}^{-1}} \right)^{-4/3} \left(\frac{D}{\rm{5 \; kpc}} \right)^{-2} \; \mu\rm{Jy}
\end{split}
\end{equation}
using the observing frequency $\nu$, the electron temperature $T_e$, the wind mass loss rate $\dot{M}$, the mean atomic weight per electron $\mu_e$, the terminal wind velocity $v_{\infty}$ and the distance $D$. For the red giant donor star in RS Oph, the terminal wind velocity is around 20 km/s and the mass loss rate is around $5 \cdot 10^{-8} \; M_{\odot} \rm{yr}^{-1}$, see \cite{walder20083d} and Section \ref{sec:lc_modelling2}. We assume an electron temperature of $10^4$ K and a mean atomic weight per electron of $\mu_e =1$. Note that the dependence on the electron temperature is very weak. Figure \ref{fig:stellar_wind} shows the stellar wind radio flux for the aforementioned parameters in the thick dashed black line. The coloured lines show the radio flux for either an increased mass loss rate (blue), distance (red) or terminal wind velocity (green). For the mass loss rate and terminal wind velocity, we multiply the initial values by a factor of two, for the distance we choose the upper limit of the Gaia DR3 distance estimate.

Based on Figure \ref{fig:stellar_wind} we conclude that the stellar wind contributes around $10-50 \; \rm{\mu}$Jy at frequencies of 100 MHz to 1 GHz. This is a negligible fraction of the flux observed during the outburst. However, if there is a quiescent radio flux of around 0.4 mJy, as suggested in Section \ref{sec:t<5_old_pop}, the stellar wind could add a measurable amount of flux to the total observed flux density.

\subsection{Radio jet} \label{sec:jet}


There are a few hints that point toward an accretion disc and jet-launching scenario in RS Oph. Although the most likely scenario for the recurrent nova phenomenon is a thermonuclear runaway on the surface of the white dwarf \citep{starrfield2008rs}, some authors suggest a different origin, more similar to dwarf novae \citep{king2009rs, alexander2011disc}. In dwarf novae, the brightening of the system originates from an accretion disc instability. A supporting observation of a disc scenario is the cataclysmic-variable-like rapid optical variability that was observed for RS Oph \citep{zamanov2018recurrent, munari2022flickering}, which is thought to originate from the inner regions of an accretion disc or from changes in the column density of the absorbing wind envelope. Furthermore, a photoionization model of the quiescent spectrum indicates the presence of a low-luminosity accretion disc \citep{mondal2018optical}. Additionally, an extremely collimated thermal jet has been observed in the 2006 outburst \citep{sokoloski2008uncovering} at 43 GHz.  This jet might also be able to generate non-thermal synchrotron emission. Bipolar outflows have also been suggested for other novae with red giant companions \citep{linford2017peculiar}. Finally, the spectrum, as shown in Figure \ref{fig:rsoph_spectrum}, is quite flat over several orders in frequency. This is typical for a jet spectrum \citep{blandford1979jets, markoff2001jet, moscibrodzka2013coupled}. The flat nature of the jet spectrum is due to the summation of partially self-absorbed synchrotron spectra from different regions of the jet, where the optical depth decreases with radius. This only results in a flat spectrum if the energy losses due to adiabatic decompression are neglected.

To examine whether the (flat-spectrum) radio emission observed from RS Oph originates from radio jets, we apply the \cite{blandford1979jets} jet model to our data. The \cite{blandford1979jets} model assumes a narrow conical jet with constant particle acceleration within the jet. The model predicts a maximum jet radius, which we compare to the size of the observed thermal jet. Appendix \ref{ap:jet} describes the model in more detail and shows how to derive the following equation for the maximum jet radius from \cite{blandford1979jets}. The maximum jet radius is derived in Eqn. \ref{eq:rmax}
\begin{equation}
\begin{split}
    & r_{\rm{max,obs}}  \approx 37 \left[\frac{k_e}{0.5} \right]^{-3/51} \rm{sin}(\theta) ^{3/51} \mathcal{D}_j^{-18/51} (1+z)^{-25/17}
    \\
    & \times{} \left[\frac{S_{\rm{obs}}}{5.6 \; \rm{mJy}} \right]^{8/17} \left[\frac{\phi_{\rm{obs}}}{45 ^\circ} \right]^{-9/17} \left[\frac{D_l}{2.68 \; \rm{kpc}} \right]^{16/17} \left[\frac{\nu}{1.7 \; \rm{GHz}} \right]^{-1} \; \rm{AU}
\end{split}
\label{eq:rmax_simple} 
\end{equation}

where z is the redshift, $S_{\rm{obs}}$ is the observed flux of the flat spectrum, $\phi_{\rm{obs}}$ is the observed semi-angle of the jet, $D_l$ is the luminosity distance to the jet, $k_e$ is a factor of order unity that sets the magnetic energy density, $\theta$ is the angle between the observer and the jet velocity, $\mathcal{D}_j$ is the Doppler factor and $\nu$ is the observing frequency. The Doppler factor is defined as $\mathcal{D}_j = 1/\left[\gamma (1 - \beta \rm{cos}(\theta)) \right]$. The intrinsic jet velocity can be derived from the observed jet velocity by inverting \cite{blandford1979jets} Eqn. 1:
\begin{equation}
    \beta = \frac{\beta_{\rm{obs}}}{\rm{sin}(\theta)^2 + \beta_{\rm{obs}} \rm{cos}(\theta)} .
    \label{eq:beta_beta_obs}
\end{equation}

From Eqn. \ref{eq:rmax_simple} it is clear that the lowest frequency observed to be part of the flat spectrum corresponds to the maximum jet radius.

\begin{figure}
    \centering
    \includegraphics[width=\linewidth]{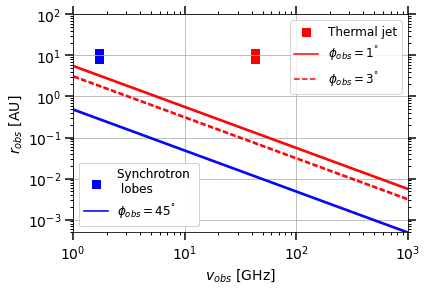}
    \caption{\protect \cite{blandford1979jets} maximum observed jet radius in astronomical units as a function of observing frequency in GHz. Here we assume $\beta~\epsilon~[0.005, 0.336]$,  $\theta =50 \degree$, $k_e = 0.5$ and $D_l = 2.68 ^{+0.17}_{-0.15}$ kpc. For the non-thermal scenario (blue) we assume $S_{\rm{obs}}=5.6$ mJy \protect\citep{munari20212021} and $\phi_{\rm{obs}} \sim 45 ^\circ$. The observed size of the synchrotron lobes is around 10 AU \protect\citep{munari20212021}. For the thermal jet scenario (red) we assume $S_{\rm{obs}} =14 $ mJy and the observed half-opening angle to be $\phi_{\rm{obs}} < 4 ^\circ$ \protect\citep{sokoloski2008uncovering}. The red square shows the size of the thermal jet as imaged by the VLA at 43 GHz \protect\citep{sokoloski2008uncovering}. }
    \label{fig:jet_radius2}
\end{figure}

As in section \ref{subsec:brightness_temp}, we assume an expansion velocity for the ejecta. However, the jet expansion velocity is not necessarily equal to the shell ejecta expansion velocity. Therefore, we gather all expansion velocities observed for various emitting components to explore the full parameter space. The radio observations of the 2006 outburst showed expansion velocities with values as low as 2400 km/s \citep{o2006asymmetric} to values as high as 11000 km/s \citep{rupen2008expanding}, which is roughly the escape velocity of a white dwarf. Furthermore, we again assume a distance of distance of $2.68 ^{+0.17}_{-0.15}$ kpc, and correct all aforementioned work to this distance. 

The angle between us and the jet, $\theta$ in Fig. \ref{fig:jet_geometry}, is unknown. Therefore, we can derive the intrinsic jet velocity, $\beta$, as a function of $\theta$ for the minimum and maximum values of $\beta_{\rm{obs}}$ using Eqn. \ref{eq:beta_beta_obs}. Based on the aforementioned observed expansion velocities and exploring a wide range of moderate $\theta$ values ($\theta~\epsilon~[15^{\circ}, 75^{\circ}]$), the intrinsic jet velocity lies between $\beta = 0.005$ and $\beta=0.336$. We point out that the observed jet radius has a very weak dependence on the jet velocity with our range of $\beta~\epsilon~[0.008, 0.037]$ and $\theta$. Dependence on the unknown $k_e$ is very weak and we use a value of 0.5 \citep{blandford1979jets}. We will use these values and boundaries for the parameters space we will consider in the following. 

Now we calculate the maximum jet radius based on Eqn. \ref{eq:rmax_simple}, as a function of observing frequency. 
We consider two scenarios to compare to observations. In one we consider the parameters for the observed radio lobes by \cite{munari20212021} at 1.7 GHz. The predicted jet radius as a function of frequency is shown in blue for this scenario in Figure \ref{fig:jet_radius2}. We assume $S_{\rm{obs}}=5.6$ mJy  and $\phi_{\rm{obs}} \sim 45 ^\circ$ \citep{munari20212021}. The observed size of the synchrotron lobes is around 4 mas, which corresponds to 10 AU \citep{munari20212021} and is indicated with the blue square in the Figure. It is clear that there is over an order of magnitude difference in the observed size of the synchrotron lobes and the predicted \cite{blandford1979jets} jet radius.

In a second scenario we compare the thermal jets discovered by \cite{sokoloski2008uncovering} at 43 GHz against the \cite{blandford1979jets} jet model. Assuming that the thermal jet structure would also be capable of particle acceleration giving rise to synchrotron emission, we explore it in the context of the \citealt{blandford1979jets} jet model. In this scenario, the jet half-opening angle has been observed to be quite narrow, $\leq 4 \degree$ \citep{sokoloski2008uncovering}. Here we consider both $\phi_{\rm{obs}}=1^\circ$ and $\phi_{\rm{obs}}=3^\circ$. We use a flux density of $S_{\rm{obs}}=14$ mJy, as the values of the highly collimated outflows were between 13 and 15 mJy \citep{sokoloski2008uncovering}. Predicted jet radii are shown in red in Figure \ref{fig:jet_radius2}. The observed size of the thermal jet is similar to the lobes in the previous scenario; around 10 AU. Again, we find a large discrepancy between the observed thermal jet size and the predicted synchrotron jet size (assuming the \cite{blandford1979jets} scenario). Flipping the argument around, a flux density about three orders of magnitude higher than observed would be required to match the thermal jet size to the synchrotron jet size at 43 GHz. We, therefore, conclude that the flat spectrum of RS Oph cannot be explained by a jet model.

\section{Future observations}
\label{sec:future}

\subsection{Radio observation strategy}
During the next outburst of RS Oph it would be critical to start low frequency (<$1$ GHz)  monitoring early (t$\sim3$ days) after outburst. This allows one to model the rise of the synchrotron emission component at low frequencies, possibly even observing the synchrotron self-absorption frequency moving to lower and lower frequencies over time. Early observations could possibly also allow one to determine the synchrotron self-absorption frequency. Furthermore, follow-up observations long after the outburst would be interesting to probe whether and when the low-frequency radio emission drops below the detection threshold between outbursts. The inverted spectrum around day $3$ shows that either a non-thermal component rises early on but gets overwhelmed by the thermal component at higher frequencies, or a non-thermal (synchrotron) component from the previous outburst was still visible at that time. 

In order to fully model the emission components that contribute to the radio emission in RS Oph, a detailed high spatial resolution monitoring campaign is necessary. For example, in \cite{rupen2008expanding} the spectral index could be determined separately for the central component and the lobes. The angular extent of the radio structure was observed to be around $90$ mas at day $20$ by \cite{munari2022radio}. It is currently not possible to resolve this small structure with LOFAR. Including the LOFAR international baselines allows for a resolution of $300$ mas \citep{morabito2022sub}. Even the SKA-MID, with a resolution of $30$ mas, could only resolve this structure down to a few pixels \citep{SKA2022McMullin}. The $2021$ outburst of RS Oph was observed with the EVN by \cite{munari2022radio}, however, only one frequency was observed at the time, which yields pseudo-matching temporal data; not allowing for a spectral index measurement of the various components. To capture the rise and evolution of all emission components it would be necessary to observe the outburst at high spatial resolution at multiple frequencies.

\subsection{Multi-wavelength observation strategy}
 
Detailed monitoring of the optical magnitude evolution of the $2006$ and $2021$ outbursts has shown an almost identical evolution. This indicates that the white dwarf mass, ejecta mass, ejecta velocity and geometry are similar since optical evolution would be affected by these parameters \citep{shore2012spectroscopy, chomiuk2021new}. X-ray monitoring has shown that the evolution of the [Fe x] coronal emission line emission appears to be the same for both eruptions \citep{page2022swift}. Furthermore, X-ray monitoring shows that the harder X-ray emission, originating from shock interactions of the nova ejecta and the red giant wind, was similar between the two most recent eruptions. This, just as the optical evolution, implies similar ejecta parameters. However, the soft X-ray emission was found to be much brighter in the $2006$ eruption compared to the $2021$ eruption. This difference is not simply explained by higher absorption in the $2021$ eruption \citep{page2022swift}. \cite{cheung2022fermi} describe the first gamma-ray detection of RS Oph outburst, allowing them to constrain the maximum energy of the accelerated protons. Detailed gamma-ray light curves could help probe the density fluctuations of the swept-up material by the shock in the inner part of the binary and/ or changes in the outflow. Detections and upper limits of the neutrino flux can help constrain the possible hadronic origin of the gamma-ray emission \citep{ATEL_icecube, guetta2022neutrino}.

Combining the detailed radio observations at earlier and later epochs with observations at multi-wavelengths will help unravel the physical details of RS Ophiuchi. Modelling at a broad range of wavelengths will allow us to get a better idea of the density (profile) of the circumbinary medium, seed electron population, the ejecta mass, ejecta velocity, white dwarf mass, jetted or lobe-like emission components and magnetic field strengths.

\section{Conclusions} 
\label{sec:conclusion}

The symbiotic recurrent nova RS Oph was observed with LOFAR and MeerKAT for $7$ months during the $2021$ outburst. The radio emission was first detected with MeerKAT $2$ days after the optical discovery. The radio emission peaks after $15$ days at MeerKAT frequencies. The nova was also detected with LOFAR, revealing a light curve that peaks at later times compared to MeerKAT. 
The light curve can, to first order, be modelled by a single component radio supernova model, however, the data favour a second synchrotron component. The plateauing or double-peaked behaviour of the light curve can be explained by a non-uniformity in the circumbinary medium or by an additional emission component. The modelling is not able to capture the radio flux at MeerKAT frequencies at early times ($t<5$ days), however, we show that these flux levels could be explained by an old synchrotron emitting component from the previous outburst. The negative spectral index at day 3 also hints towards this scenario.
The spectrum points to multiple thermal and non-thermal emission components, which is supported by long baseline imaging \citep{munari20212021} and evidence from previous outbursts \citep{rupen2008expanding, sokoloski2006x, o2006asymmetric}. Assuming equipartition between the energy in electrons and energy in the magnetic field, we can estimate the equipartition magnetic field. For a range of ejecta expansion velocities, between $10^3-10^4$ km/s the equipartition magnetic field is between 5 and 50 mG. Finally, we show that the observed radio emission during the outburst cannot be explained by the red giant stellar wind or a synchrotron radio jet. 
A future outburst of RS Oph is expected around 2036. During this future outburst, we urge for early low-frequency observations to probe the rise of the low-frequency emission component. This could confirm whether or not the low-frequency part of the spectrum is dominated by synchrotron emission. Additionally, monitoring during optical quiescence can confirm whether and when the radio emission drops below the detection threshold between outbursts, or whether there is remnant synchrotron emission from a population of electrons.

\section*{Acknowledgements}

The authors thank the anonymous referee for constructive comments that improved the quality of this work. We would also like to thank the science directors of LOFAR and MeerKAT for their support of the Target of Opportunity observations, and the SARAO, ASTRON, MRAO and JBO staff for carrying out these observations. 

We acknowledge with thanks the variable star observations from the AAVSO International Database contributed by observers worldwide and used in this research.

The MeerKAT telescope is operated by the South African Radio Astronomy Observatory, which is a facility of the National Research Foundation, an agency of the Department of Science and Innovation. 

This paper is based (in part) on data obtained with the International LOFAR Telescope (ILT) under project codes (\texttt{DDT16\_001} and \texttt{DDT17\_002}). LOFAR \citep{van2013lofar} is the Low Frequency Array designed and constructed by ASTRON. It has observing, data processing, and data storage facilities in several countries, which are owned by various parties (each with their own funding sources), and which are collectively operated by the ILT foundation under a joint scientific policy. The ILT resources have benefited from the following recent major funding sources: CNRS-INSU, Observatoire de Paris and Universit\'{e} d’Orl\'{e}ans, France; BMBF, MIWF-NRW, MPG, Germany; Science Foundation Ireland (SFI), Department of Business, Enterprise and Innovation (DBEI), Ireland; NWO, The Netherlands; The Science and Technology Facilities Council, UK; Ministry of Science and Higher Education, Poland. 
AMI-LA is supported by the Universities of Cambridge and Oxford. \textit{e}-MERLIN is a National Facility operated on behalf of the Science \& Technology Facilities Council by the University of Manchester at Jodrell Bank Observatory.

This publication is part of the project CORTEX (NWA.1160.18.316) of the research programme NWA-ORC which is (partly) financed by the Dutch Research Council (NWO).

PAW acknowledges financial support from the University of Cape Town and the National Research Foundation (grant number: 129359). MMN acknowledges the financial support in part by the National Science Foundation under Grant No. NSF PHY-1748958.

\section*{Data Availability}

The data and data products presented in this paper are available in a reproduction package via Zenodo, at \url{https://dx.doi.org/10.5281/zenodo.7540842}.

LOFAR data for RS Oph are available via the LOFAR Long Term Archive (LTA; \url{https://lta.lofar.eu/}) under project codes \texttt{DDT16\_001} and \texttt{DDT17\_002}.
MeerKAT data for RS Oph are available from the MeerKAT archive, \url{https://archive.sarao.ac.za/} under the project name DDT-20210810-MN-01.




\bibliographystyle{mnras}
\bibliography{ms} 

\begin{thebibliography}{}
\makeatletter
\relax
\def\mn@urlcharsother{\let\do\@makeother \do\$\do\&\do\#\do\^\do\_\do\%\do\~}
\def\mn@doi{\begingroup\mn@urlcharsother \@ifnextchar [ {\mn@doi@}
  {\mn@doi@[]}}
\def\mn@doi@[#1]#2{\def\@tempa{#1}\ifx\@tempa\@empty \href
  {http://dx.doi.org/#2} {doi:#2}\else \href {http://dx.doi.org/#2} {#1}\fi
  \endgroup}
\def\mn@eprint#1#2{\mn@eprint@#1:#2::\@nil}
\def\mn@eprint@arXiv#1{\href {http://arxiv.org/abs/#1} {{\tt arXiv:#1}}}
\def\mn@eprint@dblp#1{\href {http://dblp.uni-trier.de/rec/bibtex/#1.xml}
  {dblp:#1}}
\def\mn@eprint@#1:#2:#3:#4\@nil{\def\@tempa {#1}\def\@tempb {#2}\def\@tempc
  {#3}\ifx \@tempc \@empty \let \@tempc \@tempb \let \@tempb \@tempa \fi \ifx
  \@tempb \@empty \def\@tempb {arXiv}\fi \@ifundefined
  {mn@eprint@\@tempb}{\@tempb:\@tempc}{\expandafter \expandafter \csname
  mn@eprint@\@tempb\endcsname \expandafter{\@tempc}}}

\bibitem[\protect\citeauthoryear{Acciari et~al.,}{Acciari
  et~al.}{2022}]{acciari2022gamma}
Acciari V.,  et~al., 2022, arXiv preprint arXiv:2202.07681

\bibitem[\protect\citeauthoryear{Alexander, Wynn, King  \& Pringle}{Alexander
  et~al.}{2011}]{alexander2011disc}
Alexander R.,  Wynn G.,  King A.,   Pringle J.,  2011, \mn@doi [\mnras]
  {10.1111/j.1365-2966.2011.19647.x}, \href
  {https://ui.adsabs.harvard.edu/abs/2011MNRAS.418.2576A} {418, 2576}

\bibitem[\protect\citeauthoryear{Anupama}{Anupama}{2008}]{anupama2008recurrent}
Anupama G.,  2008, in RS Ophiuchi (2006) and the Recurrent Nova Phenomenon.
  p.~31

\bibitem[\protect\citeauthoryear{Barry, Mukai, Sokoloski, Danchi, Hachisu,
  Evans, Gehrz  \& Mikolajewska}{Barry et~al.}{2008}]{barry2008distance}
Barry R.,  Mukai K.,  Sokoloski J.,  Danchi W.,  Hachisu I.,  Evans A.,  Gehrz
  R.,   Mikolajewska J.,  2008, in RS Ophiuchi (2006) and the Recurrent Nova
  Phenomenon. p.~52

\bibitem[\protect\citeauthoryear{Beck \& Krause}{Beck \&
  Krause}{2005}]{beck2005revised}
Beck R.,  Krause M.,  2005, \mn@doi [Astronomische Nachrichten]
  {10.1002/asna.200510366}, \href
  {https://ui.adsabs.harvard.edu/abs/2005AN....326..414B} {326, 414}

\bibitem[\protect\citeauthoryear{Blandford \& K{\"o}nigl}{Blandford \&
  K{\"o}nigl}{1979}]{blandford1979jets}
Blandford R.,  K{\"o}nigl A.,  1979, \mn@doi [\apj] {10.1086/157262}, 232, 34

\bibitem[\protect\citeauthoryear{Bode \& Evans}{Bode \&
  Evans}{2008}]{bode2008classical}
Bode M.~F.,  Evans A.,  2008, Classical Novae, \href
  {https://ui.adsabs.harvard.edu/abs/2008clno.book.....B} {43}

\bibitem[\protect\citeauthoryear{Bode \& Kahn}{Bode \&
  Kahn}{1985}]{bode1985model}
Bode M.,  Kahn F.,  1985, \mn@doi [\mnras] {10.1093/mnras/217.1.205}, \href
  {https://ui.adsabs.harvard.edu/abs/1985MNRAS.217..205B} {217, 205}

\bibitem[\protect\citeauthoryear{Booth, Mohamed  \& Podsiadlowski}{Booth
  et~al.}{2016}]{booth2016modelling}
Booth R.~A.,  Mohamed S.,   Podsiadlowski P.,  2016, \mn@doi [\mnras]
  {10.1093/mnras/stw001}, \href
  {https://ui.adsabs.harvard.edu/abs/2016MNRAS.457..822B} {457, 822}

\bibitem[\protect\citeauthoryear{{Brandi}, {Quiroga}, {Miko{\l}ajewska},
  {Ferrer}  \& {Garc{\'\i}a}}{{Brandi} et~al.}{2009}]{brandi2009spectroscopic}
{Brandi} E.,  {Quiroga} C.,  {Miko{\l}ajewska} J.,  {Ferrer} O.~E.,
  {Garc{\'\i}a} L.~G.,  2009, \mn@doi [\aap] {10.1051/0004-6361/200811417},
  \href {https://ui.adsabs.harvard.edu/abs/2009A&A...497..815B} {497, 815}

\bibitem[\protect\citeauthoryear{{CASA Team} et~al.,}{{CASA Team}
  et~al.}{2022}]{bean2022casa}
{CASA Team} et~al., 2022, \mn@doi [\pasp] {10.1088/1538-3873/ac9642}, \href
  {https://ui.adsabs.harvard.edu/abs/2022PASP..134k4501C} {134, 114501}

\bibitem[\protect\citeauthoryear{{Carbone} et~al.,}{{Carbone}
  et~al.}{2018}]{carbone2018pyse}
{Carbone} D.,  et~al., 2018, \mn@doi [Astronomy and Computing]
  {10.1016/j.ascom.2018.02.003}, \href
  {https://ui.adsabs.harvard.edu/abs/2018A&C....23...92C} {23, 92}

\bibitem[\protect\citeauthoryear{Cheung et~al.,}{Cheung
  et~al.}{2022}]{cheung2022fermi}
Cheung C.,  et~al., 2022, \mn@doi [\apj] {10.3847/1538-4357/ac7eb7}, \href
  {https://ui.adsabs.harvard.edu/abs/2022ApJ...935...44C} {935, 44}

\bibitem[\protect\citeauthoryear{Chevalier}{Chevalier}{1982a}]{chevalier1982radio}
Chevalier R.,  1982a, \mn@doi [\apj] {10.1086/160167}, \href
  {https://ui.adsabs.harvard.edu/abs/1982ApJ...259..302C} {259, 302}

\bibitem[\protect\citeauthoryear{Chevalier}{Chevalier}{1982b}]{chevalier1982young}
Chevalier R.,  1982b, \mn@doi [\apj] {10.1086/183853}, \href
  {https://ui.adsabs.harvard.edu/abs/1982ApJ...259L..85C} {259, L85}

\bibitem[\protect\citeauthoryear{{Chevalier}}{{Chevalier}}{1998}]{chevalier1998synchrotron}
{Chevalier} R.~A.,  1998, \mn@doi [\apj] {10.1086/305676}, \href
  {https://ui.adsabs.harvard.edu/abs/1998ApJ...499..810C} {499, 810}

\bibitem[\protect\citeauthoryear{Chomiuk et~al.,}{Chomiuk
  et~al.}{2014}]{chomiuk2014binary}
Chomiuk L.,  et~al., 2014, Nature, 514, 339

\bibitem[\protect\citeauthoryear{{Chomiuk}, {Metzger}  \& {Shen}}{{Chomiuk}
  et~al.}{2021}]{chomiuk2021new}
{Chomiuk} L.,  {Metzger} B.~D.,   {Shen} K.~J.,  2021, \mn@doi [\araa]
  {10.1146/annurev-astro-112420-114502}, \href
  {https://ui.adsabs.harvard.edu/abs/2021ARA&A..59..391C} {59, 391}

\bibitem[\protect\citeauthoryear{Collaboration*† et~al.,}{Collaboration*†
  et~al.}{2022}]{hess2022time}
Collaboration*† H.,  et~al., 2022, \mn@doi [Science]
  {10.1126/science.abn0567}, \href
  {https://ui.adsabs.harvard.edu/abs/2022Sci...376...77A} {376, 77}

\bibitem[\protect\citeauthoryear{{Darnley}}{{Darnley}}{2021}]{darnley2021accrete}
{Darnley} M.~J.,  2021, in The Golden Age of Cataclysmic Variables and Related
  Objects V. p.~44 (\mn@eprint {arXiv} {1912.13209}),
  \mn@doi{10.22323/1.368.0044}

\bibitem[\protect\citeauthoryear{{Das} \& {Mondal}}{{Das} \&
  {Mondal}}{2015}]{das2015abundance}
{Das} R.,  {Mondal} A.,  2015, \mn@doi [\na] {10.1016/j.newast.2015.02.004},
  \href {https://ui.adsabs.harvard.edu/abs/2015NewA...39...19D} {39, 19}

\bibitem[\protect\citeauthoryear{De~Gasperin et~al.,}{De~Gasperin
  et~al.}{2019}]{de2019systematic}
De~Gasperin F.,  et~al., 2019, \mn@doi [\aap] {10.1051/0004-6361/201833867},
  622, A5

\bibitem[\protect\citeauthoryear{De~Gasperin et~al.,}{De~Gasperin
  et~al.}{2020}]{de2020reaching}
De~Gasperin F.,  et~al., 2020, \mn@doi [\aap] {10.1051/0004-6361/202038663},
  642, A85

\bibitem[\protect\citeauthoryear{{Diesing}, {Metzger}, {Aydi}, {Chomiuk},
  {Vurm}, {Gupta}  \& {Caprioli}}{{Diesing} et~al.}{2022}]{diesing2022evidence}
{Diesing} R.,  {Metzger} B.~D.,  {Aydi} E.,  {Chomiuk} L.,  {Vurm} I.,  {Gupta}
  S.,   {Caprioli} D.,  2022, arXiv e-prints, \href
  {https://ui.adsabs.harvard.edu/abs/2022arXiv221102059D} {p. arXiv:2211.02059}

\bibitem[\protect\citeauthoryear{Dobrzycka \& Kenyon}{Dobrzycka \&
  Kenyon}{1994}]{dobrzycka1994new}
Dobrzycka D.,  Kenyon S.~J.,  1994, \mn@doi [\aj] {10.1086/117238}, \href
  {https://ui.adsabs.harvard.edu/abs/1994AJ....108.2259D} {108, 2259}

\bibitem[\protect\citeauthoryear{{Enoto} et~al.,}{{Enoto}
  et~al.}{2021}]{ATEL_nicer}
{Enoto} T.,  et~al., 2021, The Astronomer's Telegram, \href
  {https://ui.adsabs.harvard.edu/abs/2021ATel14850....1E} {14850, 1}

\bibitem[\protect\citeauthoryear{Eyres et~al.,}{Eyres
  et~al.}{2009}]{eyres2009double}
Eyres S.,  et~al., 2009, \mn@doi [\mnras] {10.1111/j.1365-2966.2009.14633.x},
  \href {https://ui.adsabs.harvard.edu/abs/2009MNRAS.395.1533E} {395, 1533}

\bibitem[\protect\citeauthoryear{{Fajrin}, {Imaduddin}, {Malasan}, {Arai},
  {Shinnaka}  \& {Kawakita}}{{Fajrin} et~al.}{2021}]{fajrin2021atel}
{Fajrin} M.,  {Imaduddin} I.,  {Malasan} H.~L.,  {Arai} A.,  {Shinnaka} Y.,
  {Kawakita} H.,  2021, The Astronomer's Telegram, \href
  {https://ui.adsabs.harvard.edu/abs/2021ATel14909....1F} {14909, 1}

\bibitem[\protect\citeauthoryear{Fekel, Joyce, Hinkle  \& Skrutskie}{Fekel
  et~al.}{2000}]{fekel2000infrared}
Fekel F.~C.,  Joyce R.~R.,  Hinkle K.~H.,   Skrutskie M.~F.,  2000, \mn@doi
  [\aj] {10.1086/301260}, \href
  {https://ui.adsabs.harvard.edu/abs/2000AJ....119.1375F} {119, 1375}

\bibitem[\protect\citeauthoryear{{Ferrigno} et~al.,}{{Ferrigno}
  et~al.}{2021}]{ATEL_integral}
{Ferrigno} C.,  et~al., 2021, The Astronomer's Telegram, \href
  {https://ui.adsabs.harvard.edu/abs/2021ATel14855....1F} {14855, 1}

\bibitem[\protect\citeauthoryear{{Guetta}, {Hillman}  \& {Della
  Valle}}{{Guetta} et~al.}{2022}]{guetta2022neutrino}
{Guetta} D.,  {Hillman} Y.,   {Della Valle} M.,  2022, arXiv e-prints, \href
  {https://ui.adsabs.harvard.edu/abs/2022arXiv220904873G} {p. arXiv:2209.04873}

\bibitem[\protect\citeauthoryear{Hachisu \& Kato}{Hachisu \&
  Kato}{2001}]{hachisu2001recurrent}
Hachisu I.,  Kato M.,  2001, \mn@doi [\apj] {10.1086/321601}, \href
  {https://ui.adsabs.harvard.edu/abs/2001ApJ...558..323H} {558, 323}

\bibitem[\protect\citeauthoryear{{Heywood}}{{Heywood}}{2020}]{heywood2020oxkat}
{Heywood} I.,  2020, {oxkat: Semi-automated imaging of MeerKAT observations},
  Astrophysics Source Code Library, record ascl:2009.003 (\mn@eprint {ascl}
  {2009.003})

\bibitem[\protect\citeauthoryear{Hjellming, Van~Gorkom, Taylor, Sequist, Padin,
  Davis  \& Bode}{Hjellming et~al.}{1986}]{hjellming1986radio}
Hjellming R.,  Van~Gorkom J.,  Taylor A.,  Sequist E.,  Padin S.,  Davis R.,
  Bode M.,  1986, \mn@doi [\apj] {10.1086/184687}, 305, L71

\bibitem[\protect\citeauthoryear{Jonas}{Jonas}{2009}]{jonas2009meerkat}
Jonas J.~L.,  2009, \mn@doi [Proceedings of the IEEE]
  {10.1109/JPROC.2009.2020713}, 97, 1522

\bibitem[\protect\citeauthoryear{Kafka}{Kafka}{2016}]{kafka2016observations}
Kafka S.,  2016, Observations from the AAVSO International Database

\bibitem[\protect\citeauthoryear{Kantharia, Anupama, Prabhu, Ramya, Bode, Eyres
   \& O’Brien}{Kantharia et~al.}{2007}]{kantharia2007giant}
Kantharia N.,  Anupama G.,  Prabhu T.,  Ramya S.,  Bode M.,  Eyres S.,
  O’Brien T.,  2007, \mn@doi [\apj] {10.1086/522201}, \href
  {https://ui.adsabs.harvard.edu/abs/2007ApJ...667L.171K} {667, L171}

\bibitem[\protect\citeauthoryear{Kantharia et~al.,}{Kantharia
  et~al.}{2015}]{kantharia2015insights}
Kantharia N.,  et~al., 2015, \mn@doi [\mnras] {10.1093/mnrasl/slv154}, \href
  {https://ui.adsabs.harvard.edu/abs/2016MNRAS.456L..49K} {456, L49}

\bibitem[\protect\citeauthoryear{King \& Pringle}{King \&
  Pringle}{2009}]{king2009rs}
King A.,  Pringle J.,  2009, \mn@doi [\mnras]
  {10.1111/j.1745-3933.2009.00682.x}, \href
  {https://ui.adsabs.harvard.edu/abs/2009MNRAS.397L..51K} {397, L51}

\bibitem[\protect\citeauthoryear{Linford et~al.,}{Linford
  et~al.}{2017}]{linford2017peculiar}
Linford J.,  et~al., 2017, \mn@doi [\apj] {10.3847/1538-4357/aa7512}, \href
  {https://ui.adsabs.harvard.edu/abs/2017ApJ...842...73L} {842, 73}

\bibitem[\protect\citeauthoryear{Linford et~al.,}{Linford
  et~al.}{2019}]{linford2019t}
Linford J.~D.,  et~al., 2019, The Astrophysical Journal, 884, 8

\bibitem[\protect\citeauthoryear{{Longair}}{{Longair}}{2011}]{longair2010high}
{Longair} M.~S.,  2011, {High Energy Astrophysics}

\bibitem[\protect\citeauthoryear{Markoff, Falcke  \& Fender}{Markoff
  et~al.}{2001}]{markoff2001jet}
Markoff S.,  Falcke H.,   Fender R.,  2001, \mn@doi [\aap]
  {10.1051/0004-6361:20010420}, \href
  {https://ui.adsabs.harvard.edu/abs/2001A&A...372L..25M} {372, L25}

\bibitem[\protect\citeauthoryear{{McMullin} et~al.,}{{McMullin}
  et~al.}{2022}]{SKA2022McMullin}
{McMullin} J.,  et~al., 2022, in {Marshall} H.~K.,  {Spyromilio} J.,   {Usuda}
  T.,  eds,  Society of Photo-Optical Instrumentation Engineers (SPIE)
  Conference Series Vol. 12182, Ground-based and Airborne Telescopes IX. p.
  121820Q, \mn@doi{10.1117/12.2642184}

\bibitem[\protect\citeauthoryear{Mondal, Anupama, Kamath, Das, Selvakumar  \&
  Mondal}{Mondal et~al.}{2018}]{mondal2018optical}
Mondal A.,  Anupama G.,  Kamath U.,  Das R.,  Selvakumar G.,   Mondal S.,
  2018, \mn@doi [\mnras] {10.1093/mnras/stx2988}, \href
  {https://ui.adsabs.harvard.edu/abs/2018MNRAS.474.4211M} {474, 4211}

\bibitem[\protect\citeauthoryear{Moore \& Bildsten}{Moore \&
  Bildsten}{2012}]{moore2012circumstellar}
Moore K.,  Bildsten L.,  2012, The Astrophysical Journal, 761, 182

\bibitem[\protect\citeauthoryear{Morabito et~al.,}{Morabito
  et~al.}{2022}]{morabito2022sub}
Morabito L.,  et~al., 2022, \mn@doi [\aap] {10.1051/0004-6361/202140649}, \href
  {https://ui.adsabs.harvard.edu/abs/2022A&A...658A...1M} {658, A1}

\bibitem[\protect\citeauthoryear{Mo{\'s}cibrodzka \& Falcke}{Mo{\'s}cibrodzka
  \& Falcke}{2013}]{moscibrodzka2013coupled}
Mo{\'s}cibrodzka M.,  Falcke H.,  2013, \mn@doi [\aap]
  {10.1051/0004-6361/201322692}, \href
  {https://ui.adsabs.harvard.edu/abs/2013A&A...559L...3M} {559, L3}

\bibitem[\protect\citeauthoryear{Munari \& Tabacco}{Munari \&
  Tabacco}{2022}]{munari2022flickering}
Munari U.,  Tabacco F.,  2022, \mn@doi [Research Notes of the AAS]
  {10.3847/2515-5172/ac72ae}, \href
  {https://ui.adsabs.harvard.edu/abs/2022RNAAS...6..103M} {6, 103}

\bibitem[\protect\citeauthoryear{Munari \& Valisa}{Munari \&
  Valisa}{2021a}]{munari20212021}
Munari U.,  Valisa P.,  2021a, arXiv preprint arXiv:2109.01101, \href
  {https://ui.adsabs.harvard.edu/abs/2021arXiv210901101M} {p. arXiv:2109.01101}

\bibitem[\protect\citeauthoryear{{Munari} \& {Valisa}}{{Munari} \&
  {Valisa}}{2021b}]{ATEL_vel_spec4}
{Munari} U.,  {Valisa} P.,  2021b, The Astronomer's Telegram, \href
  {https://ui.adsabs.harvard.edu/abs/2021ATel14840....1M} {14840, 1}

\bibitem[\protect\citeauthoryear{Munari, Giroletti, Marcote, O’Brien, Veres,
  Yang, Williams  \& Woudt}{Munari et~al.}{2022}]{munari2022radio}
Munari U.,  Giroletti M.,  Marcote B.,  O’Brien T.,  Veres P.,  Yang J.,
  Williams D.,   Woudt P.,  2022, arXiv preprint arXiv:2209.12794

\bibitem[\protect\citeauthoryear{Natta \& Panagia}{Natta \&
  Panagia}{1984}]{natta1984extinction}
Natta A.,  Panagia N.,  1984, \mn@doi [\apj] {10.1086/162681}, \href
  {https://ui.adsabs.harvard.edu/abs/1984ApJ...287..228N} {287, 228}

\bibitem[\protect\citeauthoryear{Nyamai, Linford, Allison, Chomiuk, Woudt,
  Ribeiro  \& Sarbadhicary}{Nyamai et~al.}{2023}]{nyamai2021radio}
Nyamai M.~M.,  Linford J.~D.,  Allison J.~R.,  Chomiuk L.,  Woudt P.~A.,
  Ribeiro V. A. R.~M.,   Sarbadhicary S.~K.,  2023, Synchrotron emission from
  double-peaked radio light curves of the symbiotic recurrent nova V3890
  Sagitarii, \mn@doi{10.48550/ARXIV.2301.09116}, \url
  {https://arxiv.org/abs/2301.09116}

\bibitem[\protect\citeauthoryear{O'Brien et~al.,}{O'Brien
  et~al.}{2006}]{o2006asymmetric}
O'Brien T.,  et~al., 2006, \mn@doi [Nature] {10.1038/nature04949}, 442, 279

\bibitem[\protect\citeauthoryear{Offringa, De~Bruyn, Biehl, Zaroubi, Bernardi
  \& Pandey}{Offringa et~al.}{2010}]{offringa2010post}
Offringa A.,  De~Bruyn A.,  Biehl M.,  Zaroubi S.,  Bernardi G.,   Pandey V.,
  2010, \mn@doi [\mnras] {10.1111/j.1365-2966.2010.16471.x}, 405, 155

\bibitem[\protect\citeauthoryear{Offringa, Van De~Gronde  \& Roerdink}{Offringa
  et~al.}{2012}]{offringa2012morphological}
Offringa A.,  Van De~Gronde J.,   Roerdink J.,  2012, \aap, \href
  {https://ui.adsabs.harvard.edu/abs/2010MNRAS.405..155O} {539, A95}

\bibitem[\protect\citeauthoryear{Offringa et~al.,}{Offringa
  et~al.}{2014}]{offringa2014wsclean}
Offringa A.,  et~al., 2014, \mn@doi [\mnras] {10.1093/mnras/stu1368}, \href
  {https://ui.adsabs.harvard.edu/abs/2014MNRAS.444..606O} {444, 606}

\bibitem[\protect\citeauthoryear{Oppenheimer \& Mattei}{Oppenheimer \&
  Mattei}{1993}]{oppenheimer1993analysis}
Oppenheimer B.~D.,  Mattei J.~A.,  1993, Journal of the American Association of
  Variable Star Observers (JAAVSO), \href
  {https://ui.adsabs.harvard.edu/abs/1993JAVSO..22..105O} {22, 105}

\bibitem[\protect\citeauthoryear{{Page} et~al.,}{{Page}
  et~al.}{2022}]{page2022swift}
{Page} K.~L.,  et~al., 2022, arXiv e-prints, \href
  {https://ui.adsabs.harvard.edu/abs/2022arXiv220503232P} {p. arXiv:2205.03232}

\bibitem[\protect\citeauthoryear{{Pandey}, {Habtie}, {Bandyopadhyay}, {Das},
  {Teyssier}  \& {Guarro Fl{\'o}}}{{Pandey} et~al.}{2022}]{pandey2022study}
{Pandey} R.,  {Habtie} G.~R.,  {Bandyopadhyay} R.,  {Das} R.,  {Teyssier} F.,
  {Guarro Fl{\'o}} J.,  2022, \mn@doi [\mnras] {10.1093/mnras/stac2079}, \href
  {https://ui.adsabs.harvard.edu/abs/2022MNRAS.515.4655P} {515, 4655}

\bibitem[\protect\citeauthoryear{{Peters}, {Clarke}, {Giacintucci}, {Kassim}
  \& {Polisensky}}{{Peters} et~al.}{2021}]{ATEL_vlite}
{Peters} W.~M.,  {Clarke} T.~E.,  {Giacintucci} S.,  {Kassim} N.~E.,
  {Polisensky} E.,  2021, The Astronomer's Telegram, \href
  {https://ui.adsabs.harvard.edu/abs/2021ATel14908....1P} {14908, 1}

\bibitem[\protect\citeauthoryear{{Pizzuto}, {Vandenbroucke}, {Santander}  \&
  {IceCube Collaboration}}{{Pizzuto} et~al.}{2021}]{ATEL_icecube}
{Pizzuto} A.,  {Vandenbroucke} J.,  {Santander} M.,   {IceCube Collaboration}
  2021, The Astronomer's Telegram, \href
  {https://ui.adsabs.harvard.edu/abs/2021ATel14851....1P} {14851, 1}

\bibitem[\protect\citeauthoryear{{Ribeiro} et~al.,}{{Ribeiro}
  et~al.}{2009}]{ribeiro2009expanding}
{Ribeiro} V.~A.~R.~M.,  et~al., 2009, \mn@doi [\apj]
  {10.1088/0004-637X/703/2/1955}, \href
  {https://ui.adsabs.harvard.edu/abs/2009ApJ...703.1955R} {703, 1955}

\bibitem[\protect\citeauthoryear{Rupen, Mioduszewski  \& Sokoloski}{Rupen
  et~al.}{2008}]{rupen2008expanding}
Rupen M.~P.,  Mioduszewski A.~J.,   Sokoloski J.~L.,  2008, \mn@doi [\apj]
  {10.1086/525555}, 688, 559

\bibitem[\protect\citeauthoryear{Schaefer}{Schaefer}{2004}]{schaefer2004rs}
Schaefer B.,  2004, \iaucirc, \href
  {https://ui.adsabs.harvard.edu/abs/2004IAUC.8396....2S} {8396, 2}

\bibitem[\protect\citeauthoryear{Schaefer}{Schaefer}{2009}]{schaefer2009orbital}
Schaefer B.~E.,  2009, \mn@doi [\apj] {10.1088/0004-637X/697/1/721}, \href
  {https://ui.adsabs.harvard.edu/abs/2009ApJ...697..721S} {697, 721}

\bibitem[\protect\citeauthoryear{Schaefer}{Schaefer}{2010}]{schaefer2010comprehensive}
Schaefer B.~E.,  2010, \mn@doi [\apjs] {10.1088/0067-0049/187/2/275}, \href
  {https://ui.adsabs.harvard.edu/abs/2010ApJS..187..275S} {187, 275}

\bibitem[\protect\citeauthoryear{Schaefer}{Schaefer}{2022}]{schaefer2022comprehensive}
Schaefer B.~E.,  2022, \mn@doi [\mnras] {10.1093/mnras/stac2900}, \href
  {https://ui.adsabs.harvard.edu/abs/2022MNRAS.tmp.2690S} {}

\bibitem[\protect\citeauthoryear{Shimwell et~al.,}{Shimwell
  et~al.}{2022}]{shimwell2022lofar}
Shimwell T.,  et~al., 2022, \mn@doi [\aap] {10.1051/0004-6361/202142484}, \href
  {https://ui.adsabs.harvard.edu/abs/2022A&A...659A...1S} {659, A1}

\bibitem[\protect\citeauthoryear{{Shore}}{{Shore}}{2012}]{shore2012spectroscopy}
{Shore} S.~N.,  2012, Bulletin of the Astronomical Society of India, \href
  {https://ui.adsabs.harvard.edu/abs/2012BASI...40..185S} {40, 185}

\bibitem[\protect\citeauthoryear{{Shore} et~al.,}{{Shore}
  et~al.}{2021}]{ATEL_vel_spec}
{Shore} S.~N.,  et~al., 2021, The Astronomer's Telegram, \href
  {https://ui.adsabs.harvard.edu/abs/2021ATel14868....1S} {14868, 1}

\bibitem[\protect\citeauthoryear{Sokoloski, Luna, Mukai  \& Kenyon}{Sokoloski
  et~al.}{2006}]{sokoloski2006x}
Sokoloski J.,  Luna G.,  Mukai K.,   Kenyon S.~J.,  2006, \mn@doi [Nature]
  {10.1038/nature04893}, \href
  {https://ui.adsabs.harvard.edu/abs/2006Natur.442..276S} {442, 276}

\bibitem[\protect\citeauthoryear{Sokoloski, Rupen  \& Mioduszewski}{Sokoloski
  et~al.}{2008}]{sokoloski2008uncovering}
Sokoloski J.,  Rupen M.,   Mioduszewski A.,  2008, \mn@doi [\apj]
  {10.1086/592602}, 685, L137

\bibitem[\protect\citeauthoryear{{Sokolovsky} et~al.,}{{Sokolovsky}
  et~al.}{2021}]{ATEL_vla}
{Sokolovsky} K.,  et~al., 2021, The Astronomer's Telegram, \href
  {https://ui.adsabs.harvard.edu/abs/2021ATel14886....1S} {14886, 1}

\bibitem[\protect\citeauthoryear{Starrfield}{Starrfield}{2008}]{starrfield2008rs}
Starrfield S.,  2008, in RS Ophiuchi (2006) and the Recurrent Nova Phenomenon.
  p.~4

\bibitem[\protect\citeauthoryear{Starrfield, Sparks  \& Truran}{Starrfield
  et~al.}{1985}]{starrfield1985recurrent}
Starrfield S.,  Sparks W.~M.,   Truran J.~W.,  1985, \apj, 291, 136

\bibitem[\protect\citeauthoryear{{Taguchi}, {Ueta}  \& {Isogai}}{{Taguchi}
  et~al.}{2021a}]{taguchi2021atel}
{Taguchi} K.,  {Ueta} T.,   {Isogai} K.,  2021a, The Astronomer's Telegram,
  \href {https://ui.adsabs.harvard.edu/abs/2021ATel14838....1T} {14838, 1}

\bibitem[\protect\citeauthoryear{{Taguchi}, {Maehara}, {Isogai}, {Tampo}  \&
  {Ito}}{{Taguchi} et~al.}{2021b}]{ATEL_vel_spec3}
{Taguchi} K.,  {Maehara} H.,  {Isogai} K.,  {Tampo} Y.,   {Ito} J.,  2021b, The
  Astronomer's Telegram, \href
  {https://ui.adsabs.harvard.edu/abs/2021ATel14858....1T} {14858, 1}

\bibitem[\protect\citeauthoryear{{Tang} \& {Chevalier}}{{Tang} \&
  {Chevalier}}{2017}]{Tang2017}
{Tang} X.,  {Chevalier} R.~A.,  2017, \mn@doi [\mnras] {10.1093/mnras/stw2978},
  \href {https://ui.adsabs.harvard.edu/abs/2017MNRAS.465.3793T} {465, 3793}

\bibitem[\protect\citeauthoryear{Tasse et~al.,}{Tasse
  et~al.}{2018}]{tasse2018faceting}
Tasse C.,  et~al., 2018, \aap, \href
  {https://ui.adsabs.harvard.edu/abs/2020A&A...642A..85D} {611, A87}

\bibitem[\protect\citeauthoryear{Taylor, Davis, Porcas  \& Bode}{Taylor
  et~al.}{1989}]{taylor1989vlbi}
Taylor A.,  Davis R.,  Porcas R.,   Bode M.,  1989, \mn@doi [\mnras]
  {10.1093/mnras/237.1.81}, \href
  {https://ui.adsabs.harvard.edu/abs/1989MNRAS.237...81T} {237, 81}

\bibitem[\protect\citeauthoryear{Van~Weeren et~al.,}{Van~Weeren
  et~al.}{2016}]{van2016lofar}
Van~Weeren R.,  et~al., 2016, \mn@doi [\apjs] {10.3847/0067-0049/223/1/2}, 223,
  2

\bibitem[\protect\citeauthoryear{Walder, Folini  \& Shore}{Walder
  et~al.}{2008}]{walder20083d}
Walder R.,  Folini D.,   Shore S.~N.,  2008, \mn@doi [\aap]
  {10.1051/0004-6361:200809703}, 484, L9

\bibitem[\protect\citeauthoryear{Weiler, Van~Dyk, Sramek  \& Panagia}{Weiler
  et~al.}{1996}]{weiler1996radio}
Weiler K.~W.,  Van~Dyk S.~D.,  Sramek R.~A.,   Panagia N.,  1996, in
  International Astronomical Union Colloquium. pp 283--297

\bibitem[\protect\citeauthoryear{Weiler, Panagia, Montes  \& Sramek}{Weiler
  et~al.}{2002}]{weiler2002radio}
Weiler K.~W.,  Panagia N.,  Montes M.~J.,   Sramek R.~A.,  2002, \mn@doi
  [\araa] {10.1146/annurev.astro.40.060401.093744}, \href
  {https://ui.adsabs.harvard.edu/abs/2002ARA&A..40..387W} {40, 387}

\bibitem[\protect\citeauthoryear{{Williams}, {O'Brien}, {Woudt}, {Nyamai},
  {Green}, {Titterington}, {Fender}  \& {Sivakoff}}{{Williams}
  et~al.}{2021}]{ATEL_ami}
{Williams} D.,  {O'Brien} T.,  {Woudt} P.,  {Nyamai} M.,  {Green} D.,
  {Titterington} D.,  {Fender} R.,   {Sivakoff} G.,  2021, The Astronomer's
  Telegram, \href {https://ui.adsabs.harvard.edu/abs/2021ATel14849....1W}
  {14849, 1}

\bibitem[\protect\citeauthoryear{Wright \& Barlow}{Wright \&
  Barlow}{1975}]{wright1975radio}
Wright A.~E.,  Barlow M.~J.,  1975, \mn@doi [\mnras] {10.1093/mnras/170.1.41},
  \href {https://ui.adsabs.harvard.edu/abs/1975MNRAS.170...41W} {170, 41}

\bibitem[\protect\citeauthoryear{Wynn}{Wynn}{2008}]{wynn2008accretion}
Wynn G.,  2008, in RS Ophiuchi (2006) and the Recurrent Nova Phenomenon. p.~73

\bibitem[\protect\citeauthoryear{Zamanov et~al.,}{Zamanov
  et~al.}{2018}]{zamanov2018recurrent}
Zamanov R.,  et~al., 2018, \mn@doi [\mnras] {10.1093/mnras/sty1816}, \href
  {https://ui.adsabs.harvard.edu/abs/2018MNRAS.480.1363Z} {480, 1363}

\bibitem[\protect\citeauthoryear{de Gasperin et~al.,}{de~Gasperin
  et~al.}{2021}]{de2021lofar}
de Gasperin F.,  et~al., 2021, Astronomy \& Astrophysics, 648, A104

\bibitem[\protect\citeauthoryear{van Haarlem et~al.,}{van Haarlem
  et~al.}{2013}]{van2013lofar}
van Haarlem M.~P.,  et~al., 2013, \mn@doi [\aap] {10.1051/0004-6361/201220873},
  \href {https://ui.adsabs.harvard.edu/abs/2013A&A...556A...2V} {556, A2}

\makeatother
\end{thebibliography}




\appendix
\section{Light curve modelling}\label{ap:lightcurve_model}
The radio supernova model that is used in the light curve modelling in Section \ref{sec:lc_modelling} is described by \cite{kantharia2007giant}. The flux density evolves as a function of time and frequency
\begin{equation}
    S\;(\rm{mJy}) = K_1 \left(\frac{\nu}{1 \; \rm{GHz}} \right)^\alpha \left(\frac{t-t_0}{20 \; \rm{days}} \right)^\beta e^{-\tau_{\rm{homog}}^{\rm{CSM}}} \left(\frac{1-e^{-\tau_{\rm{clumps}}^{\rm{CSM}}}}{\tau_{\rm{clumps}}^{\rm{CSM}}} \right)
\end{equation}

where
\begin{equation}
    \tau_{\rm{homog}}^{\rm{CSM}} = K_2 \left(\frac{\nu}{1 \; \rm{GHz}} \right)^{-2.1} \left(\frac{t-t_0}{20 \; \rm{days}} \right)^\delta
\end{equation}
and 
\begin{equation}
    \tau_{\rm{clumps}}^{\rm{CSM}} = K_3 \left(\frac{\nu}{1 \; \rm{GHz}} \right)^{-2.1} \left(\frac{t-t_0}{20 \; \rm{days}} \right)^{\delta \prime}
\end{equation}

with $K_1$, $K_2$ and $K_3$ determined from fits to the data and corresponding, formally, to the flux density ($K_1$), uniform ($K_2$, $K_3$), and clumpy or filamentary ($K_3$) absorption at 1 GHz 20 days after the explosion date $t_0$. The terms $\tau_{\rm{homog}}^{\rm{CSM}}$ and $\tau_{\rm{clumps}}^{\rm{CSM}}$ describe the attenuation of local, uniform CSM and clumpy CSM that are near enough to the supernova progenitor that they are altered by the rapidly expanding supernova blastwave. The $\tau_{\rm{homog}}^{\rm{CSM}}$ absorption is produced by an ionized medium that uniformly covers the emitting source (“uniform external absorption”), and the $(1-e^{-\tau_{\rm{clumps}}^{\rm{CSM}}})(\tau_{\rm{clumps}}^{\rm{CSM}})^{-1}$ term  describes the attenuation produced by an inhomogeneous medium (“clumpy absorption”) (see \cite{natta1984extinction} for a more detailed discussion of attenuation in inhomogeneous media). All external and clumpy absorbing media are assumed to be purely thermal, singly ionized gas that absorbs via free-free (f-f) transitions with frequency dependence $\nu^{-2.1}$ in the radio. The parameters $\delta$ and $\delta \prime$ describe the time dependence of the optical depths for the local uniform and clumpy or filamentary media, respectively.

In the modelling, we decide to fit for $t_0$. The observations that are used for the light curve fitting are all corrected to the start date of the explosion as determined by \cite{munari20212021} (MJD $59434.5$). These authors derive this start date by extrapolating the optical data back to quiescence magnitude. The radio data presented in this work agrees with this start time as we find that the models favour a $t_0$ of zero.

\section{Fit parameters} \label{ap:fit_parameters}
Table \ref{tab:fit_values_one_component} shows the lower and upper bounds for each parameter in the radio supernova model as described in Appendix \ref{ap:lightcurve_model}. The final column shows that fit parameters resulting in the model that best describes the data, as shown in the left panel of Figure \ref{fig:light_curve_fits}.
\begin{table}
    \centering
    \begin{tabular}{lcc r c l}
    \hline
                        & Lower    & Upper    &  \\
        Parameter       &  bound   & bound   & \multicolumn{3}{c}{Fit value} \\ \hline
        $K_1$           & 10            & 200           & 90.5  & $\pm$ &  12.0         \\
        $K_2$           & 1e-20         & 1e-1          & $9.0\cdot10^{-9}$  & $\pm$ &  $1.3\cdot 10^{-7}$   \\
        $K_3$           & 0             & 1             & $8.2\cdot 10^{-2}$ & $\pm$ & $7.7\cdot 10^{-2}$       \\
        $\alpha$        & -1            & 1             & $-7.4\cdot 10^{-2}$ & $\pm$ & $6.6\cdot 10^{-2}$      \\
        $\beta$         & -5            & 5             & $-6.2\cdot 10^{-1}$ & $\pm$ & $9.9\cdot 10^{-2}$      \\
        $\delta$        & -50           & 50            & -15.4 & $\pm$ & 9.0           \\
        $\delta \prime$ & -5            & 0             & -4.7  & $\pm$ &  1.2          \\
        $t_0$           & 0             & 5             & $3.5\cdot 10^{-7}$  & $\pm$ &  3.6       \\ \hline
        
    \end{tabular}
    \caption{Lower and upper bounds and fit values for the parameters in the radio supernova models, as shown in the left panel of Figure \ref{fig:light_curve_fits}.}
    \label{tab:fit_values_one_component}
\end{table}

Table \ref{tab:fit_values_two_component} shows the lower and upper bounds for each parameter in the radio supernova model as described in Appendix \ref{ap:lightcurve_model}. Here, we allow for two such components, labeled a and b in the Table. The final column shows the fit parameters resulting in the model that best describes the data, as shown in the right panel of Figure \ref{fig:light_curve_fits}.

\begin{table}
    \centering
    \begin{tabular}{lcc r c l}
    \hline
                        & Lower    & Upper    &  \\
        Parameter       &  bound   & bound   & \multicolumn{3}{c}{Fit value} \\ \hline
        $K_{1,a}$           & 10            & 200           & 62.3  & $\pm$ &  25.1         \\
        $K_{2,a}$           & 1e-5         & 1e-1          & $1.0\cdot 10^{-2}$  & $\pm$ &  $7.1\cdot 10^{-2}$   \\
        $K_{3,a}$           & 0             & 1             & $8.7\cdot 10^{-2}$ & $\pm$ & $2.5\cdot 10^{-1}$       \\
        $\alpha_a$          & -1            & 1             & $-4.9\cdot 10^{-1}$ & $\pm$ & $2.8\cdot 10^{-1}$    \\
        $\beta_a$           & -5            & 5             & -1.1 & $\pm$ & $3.8\cdot 10^{-1}$      \\
        $\delta_a$          & -50           & 50            & -4.2 & $\pm$ & 5.7           \\
        $\delta_a \prime$   & -5            & 0             & -4.2  & $\pm$ &  2.1          \\
        $t_{0,a}$           & 0             & 5             & $9.6\cdot 10^{-4}$  & $\pm$ &  2.5       \\ \hline
        
        $K_{1,b}$           & 10            & 300           & 217.6  & $\pm$ &  186.5         \\
        $K_{2,b}$           & 1e-5          & 10          & $4.0\cdot 10^{-2}$  & $\pm$ &  $1.8\cdot 10^{-1}$   \\
        $K_{3,b}$           & 0             & 10             & 6.9 & $\pm$ & 6.7       \\
        $\alpha_b$          & -2            & 2             & -1.3 & $\pm$ & 0.5      \\
        $\beta_b$           & -5            & 5             & -1.3 & $\pm$ & 0.5      \\
        $\delta_b$          & -50           & 50            & 0.2 & $\pm$ & 1.9           \\
        $\delta_b \prime$   & -5            & 0             & -2.4  & $\pm$ &  0.7          \\
        $t_{0,b}$           & 0             & 10             & 6.1  & $\pm$ &  6.8      \\ \hline
    \end{tabular}
    \caption{Lower and upper bounds and fit values for the parameters in the radio supernova models, allowing for two components, as shown in the right panel of Figure \ref{fig:light_curve_fits}.}
    \label{tab:fit_values_two_component}
\end{table}

\section{Blandford-K{\"o}nigl jet model} \label{ap:jet}

\begin{figure}
    \centering
    \includegraphics[width=\linewidth]{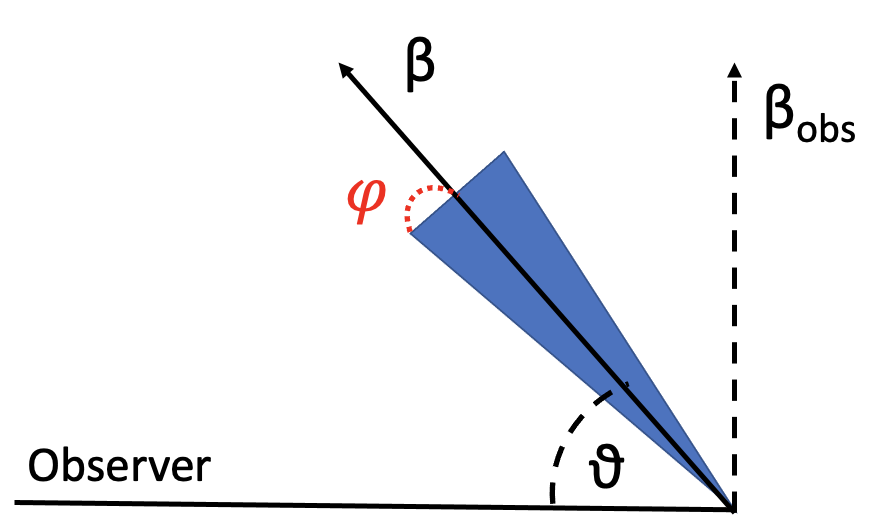}
    \caption{Schematic sketch of the jet geometry.}
    \label{fig:jet_geometry}
\end{figure}
The \cite{blandford1979jets} jet model assumes a narrow conical jet as shown in Figure \ref{fig:jet_geometry}. The jet is shown as a blue cone, with intrinsic velocity $\beta$ and semiangle $\phi$. The angle between the observer and jet velocity is $\theta$. The jet is assumed to be supersonic and free, with constant velocity $\beta$, and the magnetic field will vary as $r^{-1}$. Furthermore, we assume that relativistic electrons are continuously accelerated within the jet, and these electrons will emit synchrotron radiation with a spectral index $\alpha=-\frac{1}{2}$. The electron distribution function is $N(\gamma_e) = K \gamma_e^{-2}$ with $\gamma_{e,min} < \gamma_e < \gamma_{e,max}$. Finally, approximate equipartition between the electron energy density and magnetic energy density is assumed. 

This analysis is based on Eqn 28 and 29 from \cite{blandford1979jets}, describing the maximum jet radius in Eqn. \ref{eq:rmax} and the observed radio flux density of the flat jet spectrum in Eqn. \ref{eq:sobs}.

The observed jet spectrum is flat between $\nu_{\rm{min}}$ and $\nu_{\rm{max}}$ and the corresponding flux density is
\begin{equation}
\begin{split}
    S_{\rm{obs}} & \approx 0.5 (1+z) k_e^{5/6} \Delta^{-17/12} \left(1+ \frac{2}{3}k_e \Lambda \right)^{-17/12} \gamma^{-17/6} \\
    & \beta^{-17/12} \mathcal{D}_j^{13/6} \left[\rm{sin}(\theta)\right]^{-5/6} \phi_{\rm{obs}}^{-1} L_{44}^{17/12} D_l^{-2}
    \label{eq:sobs}
\end{split}
\end{equation}
where z is the redshift, $\gamma$ is the Lorentz factor of the jet, $\beta$ is the ratio of the jet velocity to the speed of light, $\mathcal{D}_j$ is the Doppler factor of the jet, $\phi_{\rm{obs}} = \phi \cdot \rm{csc}(\theta)$ is the observed semiangle of the jet, $L_{44} = 10^{44} L \; \rm{ergs} \; \rm{s}^-1$ is the total power carried in the jet in the form of relativistic electrons and magnetic field and $D_l$ is the luminosity distance to the jet in Gpc . Furthermore, $k_e \lesssim 1$ is a factor of order unity that sets the magnetic energy density $u_e = k_e \Lambda B^2/(8\pi)$, $\Delta = \rm{ln}(r_{max}/r_{min})$ is the ratio of the maximum and minimum jet radius, $\Lambda = \rm{ln}(\gamma_{e,max}/\gamma_{e,min})$ is the ratio of the maximum and minimum electron Lorentz factor of the seed electrons. 

By inverting Eqn \ref{eq:sobs} one can determine the total jet power $L_{44}$ 
\begin{equation}
\begin{split}
    L_{44} & \approx \Delta \left(1+ \frac{2}{3}k_e \Lambda \right)\gamma^2 \beta \\
    & \left[2 S_{\rm{obs}}^{-1} (1+z)^{-1} k_e^{-5/6} \left[\rm{sin}(\theta)\right]^{5/6} \phi_{\rm{obs}} D_l^{2} \right]^{12/17}
    \label{eq:ltot}
\end{split}
\end{equation}
which can be used to find the maximum jet radius (\cite{blandford1979jets} Eqn 28)
\begin{equation}
\begin{split}
    r_{\rm{max,obs}} & \approx 3 (1+z)^{-1} k_e^{1/3} \Delta^{-2/3} \left(1+ \frac{2}{3}k_e \Lambda \right)^{-2/3} \gamma \beta \mathcal{D}_j^{2/3}
    \\
    & \left[\rm{sin}(\theta)\right]^{-1/3} \phi_{\rm{obs}}^{-1} L_{44}^{2/3} \nu^{-1} \\
    & \approx 3 \left[2^8 (1+z)^{-25} S_{\rm{obs}}^{8} \phi_{\rm{obs}}^{-9} D_l^{16}\right]^{1/17} \\
    & \left[k_e^{-3} \left[\rm{sin}(\theta)\right]^{3} \mathcal{D}_j^{-18} \right]^{1/51} \nu^{-1} 
\end{split}
\end{equation}
here $r_{\rm{max,obs}}$ is the jet radius in pc, $S_{\rm{obs}}$ is the observed radio flux in Jy, $\phi_{\rm{obs}}$ is the observed half-opening angle in radians, $D_l$ is the luminosity distance in Gpc, $\theta$ is the angle between the observer and the jet velocity in radians and $\nu$ is the frequency in GHz. The Doppler factor is defined as $\mathcal{D}_j = 1/\left[\gamma (1 - \beta \rm{cos}(\theta)) \right]$.  Converting to a more useful notation we find:

\begin{equation}
\begin{split}
    & r_{\rm{max,obs}}  \approx 37 \left[\frac{k_e}{0.5} \right]^{-3/51} \rm{sin}(\theta) ^{3/51} \mathcal{D}_j^{-18/51} (1+z)^{-25/17}
    \\
    & \left[\frac{S_{\rm{obs}}}{5.6 \; \rm{mJy}} \right]^{8/17} \left[\frac{\phi_{\rm{obs}}}{45 ^\circ} \right]^{-9/17} \left[\frac{D_l}{2.68 \; \rm{kpc}} \right]^{16/17} \left[\frac{\nu}{1.7 \; \rm{GHz}} \right]^{-1} \; \rm{AU}
\end{split}
    \label{eq:rmax}
\end{equation}

Equation \ref{eq:rmax} is used in the analysis in Section \ref{sec:jet}.

\section{Additional data} \label{ap:additional_data}
Table \ref{tab:obs_atels_2021} shows the flux density measurements of RS Oph during the 2021 for other radio facilities. This table was put together using Astronomers Telegram's and the data is used to create the spectrum as presented in Figure \ref{fig:rsoph_spectrum}.

\begin{table}
    \centering
    \caption{Flux density values from observations of the 2021 outburst with AMI-LA, \textit{e}-MERLIN, the VLA and VLITE. The columns show the observation start date in UTC, the time since the start of the eruption in days, the observation frequency in MHz, the telescope, the observed flux density in mJy and the reference from which the data was taken. Here $t_0$ is the start of the nova eruption 2021 August 08.5 \citep{munari20212021}.\\
    References: [1]:\protect\cite{ATEL_ami},  [2]:\protect\cite{ATEL_vla}, [3]:\protect\cite{ATEL_vlite}}
    \label{tab:obs_atels_2021}

\begin{tabular}{l r@{}l r@{}l c r@{}l @{}c @{}r@{}l  r}

\hline 
  & \multicolumn{2}{c}{}   & \multicolumn{2}{c}{Obs.}  &  & \multicolumn{5}{c}{}  &  \\
Start date  & \multicolumn{2}{c}{($t-t_0$)}   & \multicolumn{2}{c}{freq.}  & Telescope & \multicolumn{5}{c}{$S_{\nu}$}  & Ref. \\

(UTC) &  \multicolumn{2}{c}{(Days)} &  \multicolumn{2}{c}{(MHz)} &   &   \multicolumn{5}{c}{(mJy)}  & \\ 
\hline
2021-08-09.77       & 1&.27        & 15&500                       & AMI-LA    & 0&.8 & $~\pm~$ & 0&.08      & [1]   \\
2021-08-10.60       & 2&.1         & 5&000                        & \textit{e}-MERLIN  & 0&.38 & $~\pm~$ & 0&.06     & [1]   \\
2021-08-10.77       & 2&.27        & 15&500                       & AMI-LA    & 1&.50 & $~\pm~$ & 0&.15     & [1]   \\
2021-08-13          & 4&.54        & 2&600                        & VLA       & 4&.71 & $~\pm~$ & 0&.14   & [2]   \\
2021-08-13          & 4&.54        & 3&400                        & VLA       & 5&.33 & $~\pm~$ & 0&.09   & [2]   \\
2021-08-13          & 4&.54        & 5&100                        & VLA       & 10&.01 & $~\pm~$ & 0&.06  & [2]   \\
2021-08-13          & 4&.54        & 7&000                        & VLA       & 15&.61 & $~\pm~$ & 0&.06  & [2]   \\
2021-08-13          & 4&.54        & 13&700                       & VLA       & 24&.09 & $~\pm~$ & 0&.06  & [2]   \\
2021-08-13          & 4&.54        & 16&500                       & VLA       & 25&.57 & $~\pm~$ & 0&.06  & [2]   \\
2021-08-13          & 4&.54        & 31&100                       & VLA       & 35&.97 & $~\pm~$ & 0&.14  & [2]   \\
2021-08-13          & 4&.54        & 34&900                       & VLA       & 38&.32 & $~\pm~$ & 0&.16  & [2]   \\
2021-08-18          & 9&.5         & &338                         & VLITE     & \textless{}12&.5 & & &  & [3] \\
2021-08-20          & 11&.5        & &338                         & VLITE     & 16&.1 & $~\pm~$ & 4&.1      & [3] \\
2021-09-04          & 26&.5        & &338                         & VLITE     & 58&.7  & $~\pm~$ & 12&.2    & [3] \\
2021-09-04          & 26&.51       & 2&600                        & VLA       & 94&.47 & $~\pm~$ & 0&.22  & [2]   \\
2021-09-04          & 26&.51       & 3&400                        & VLA       & 85&.81 & $~\pm~$ & 0&.17  & [2]   \\
2021-09-04          & 26&.51       & 5&100                        & VLA       & 65&.12 & $~\pm~$ & 0&.08  & [2]   \\
2021-09-04          & 26&.51       & 7&000                        & VLA       & 57&.62 & $~\pm~$ & 0&.06  & [2]   \\
2021-09-04          & 26&.51       & 13&700                       & VLA       & 54&.10 & $~\pm~$ & 0&.06  & [2]   \\
2021-09-04          & 26&.51       & 16&500                       & VLA       & 54&.54 & $~\pm~$ & 0&.07  & [2]   \\
2021-09-04          & 26&.51       & 31&100                       & VLA       & 70&.78 & $~\pm~$ & 0&.15  & [2]   \\
2021-09-04          & 26&.51       & 34&900                       & VLA       & 77&.04 & $~\pm~$ & 0&.16  & [2]   \\
2021-09-13          & 35&.51       & 2&600                        & VLA       & 74&.22 & $~\pm~$ & 0&.24  & [2]   \\
2021-09-13          & 35&.51       & 3&400                        & VLA       & 67&.99 & $~\pm~$ & 0&.14  & [2]   \\
2021-09-13          & 35&.51       & 5&100                        & VLA       & 56&.61 & $~\pm~$ & 0&.06  & [2]   \\
2021-09-13          & 35&.51       & 7&000                        & VLA       & 54&.47 & $~\pm~$ & 0&.06  & [2]   \\
2021-09-13          & 35&.51       & 13&700                       & VLA       & 56&.89 & $~\pm~$ & 0&.06  & [2]   \\
2021-09-13          & 35&.51       & 16&500                       & VLA       & 60&.28 & $~\pm~$ & 0&.06  & [2]   \\
2021-09-13          & 35&.51       & 31&100                       & VLA       & 92&.38 & $~\pm~$ & 0&.14  & [2]   \\
2021-09-13          & 35&.51       & 34&900                       & VLA       & 103&.52 & $~\pm~$ & 0&.15 & [2]   \\
\hline
\end{tabular}

\end{table}

Table \ref{tab:obs_2006} shows the flux density measurements of RS Oph during the 2006 outburst. Measurements from MERLIN, the VLA and the GMRT are included. The data in this table is used to create the spectrum as presented in Figure \ref{fig:rsoph_spectrum}.

\begin{table}
    \centering
    \caption{Flux density values from observations of the 2006 outburst with MERLIN, the VLA and the GMRT. The columns show the observation start date in UTC, the time since the start of the eruption in days, the observation frequency in MHz, the telescope, the observed flux density in mJy, and the reference from which the data was taken. Here $t_0$ is the start of the nova eruption 2006 February 12.8 \citep{munari20212021}.\\
    References: [4]:\protect\cite{eyres2009double}, [5]:\protect\cite{kantharia2007giant}}
    \label{tab:obs_2006}

\begin{tabular}{l r@{}l r@{}l c r@{}l c r@{}l  r}

\hline 
  & \multicolumn{2}{c}{}   & \multicolumn{2}{c}{Obs.}  &  & \multicolumn{5}{c}{}  &  \\
Start date  & \multicolumn{2}{c}{($t-t_0$)}   & \multicolumn{2}{c}{freq.}  & Telescope & \multicolumn{5}{c}{$S_{\nu}$}  & Ref. \\

(UTC) &  \multicolumn{2}{c}{(Days)} &  \multicolumn{2}{c}{(MHz)} &   &   \multicolumn{5}{c}{(mJy)}  & \\ 
\hline
2006-02-17          & 4&.5         & 6&000                        & MERLIN    & 14& & $~\pm~$ & 2&          & [4]    \\
2006-02-17          & 4&.7         & 1&460                        & VLA       & 2&.8 & $~\pm~$ & 0&.2       & [4]    \\
2006-02-17          & 4&.7         & 4&890                        & VLA       & 15&.2 & $~\pm~$ & 0&.2      & [4]    \\
2006-02-17          & 4&.7         & 14&960                       & VLA       & 23&.2 & $~\pm~$ & 0&.6      & [4]    \\
2006-02-17          & 4&.7         & 22&480                       & VLA       & 26&.3 & $~\pm~$ & 0&.5      & [4]    \\
2006-02-20          & 7&.46        & 6&000                        & MERLIN    & 41&.2 & $~\pm~$ & 0&.8      & [4]    \\
2006-03-13          & 27&.85       & 1&460                        & VLA       & 50&.4 & $~\pm~$ & 0&.4      & [4]    \\
2006-03-13          & 27&.85       & 4&890                        & VLA       & 50&.3 & $~\pm~$ & 0&.2      & [4]    \\
2006-03-13          & 27&.85       & 14&960                       & VLA       & 52&.9 & $~\pm~$ & 0&.6      & [4]    \\
2006-03-13          & 27&.85       & 22&480                       & VLA       & 69&.4 & $~\pm~$ & 0&.7      & [4]    \\
2006-03-14          & 28&.5        & 6&000                        & MERLIN    & 45& & $~\pm~$ & 5&          & [4]    \\
2006-03-15          & 29&.15       & &240                         & GMRT      & \textless 13&.0  & & &     & [5] \\
2006-03-15          & 29&.15       & &610                         & GMRT      & 48&.9 & $~\pm~$ & 7&.3      & [5] \\
2006-03-29          & 45&.07       & &240                         & GMRT      & 54&.2 & $~\pm~$ & 8&.0      & [5] \\
2006-03-29          & 45&.07       & &610                         & GMRT      & 47&.9 & $~\pm~$ & 7&.2      & [5] \\
2006-03-30          & 46&.57       & 1&460                        & VLA       & 52&.9 & $~\pm~$ & 0&.2      & [4]    \\
2006-03-30          & 46&.57       & 4&890                        & VLA       & 47&.9 & $~\pm~$ & 0&.2      & [4]    \\
2006-03-30          & 46&.57       & 14&960                       & VLA       & 48&.8 & $~\pm~$ & 0&.6      & [4]    \\
2006-03-30          & 46&.57       & 22&480                       & VLA       & 65&.4 & $~\pm~$ & 0&.6      & [4]    \\
2006-04-06          & 52&.71       & 1&460                        & VLA       & 48&.6 & $~\pm~$ & 0&.2      & [4]    \\
2006-04-06          & 52&.71       & 4&890                        & VLA       & 42&.4 & $~\pm~$ & 0&.3      & [4]    \\
2006-04-06          & 52&.71       & 14&960                       & VLA       & 50&.9 & $~\pm~$ & 0&.7      & [4]    \\
2006-04-06          & 52&.71       & 22&480                       & VLA       & 66&.0 & $~\pm~$ & 0&.1      & [4]    \\
2006-04-07          & 53&.24       & &325                         & GMRT      & 57&.0 & $~\pm~$ & 8&.6      & [5] \\
2006-11-19          & 221&.86      & &610                         & GMRT      & 4&.2 & $~\pm~$ & 2&.1       & [5] \\
2006-11-13          & 225&.61      & &325                         & GMRT      & 8&.1 & $~\pm~$ & 4&.4       & [5] \\     \hline  
\end{tabular}

\end{table}


\bsp	
\label{lastpage}
\end{document}